\documentclass{aa}  

\usepackage{graphicx}
\usepackage{subcaption}
\usepackage{txfonts}
\begin{document}

   \title{Selected aspects of the analysis of molecular line \\ observations of edge-on circumbinary disks}
   \titlerunning{Molecular line observation of edge-on circumbinary disks}

   \author{R.~Avramenko
          \inst{1}
          \and
          S.~Wolf\inst{1}
          \and
          T.F.~Illenseer\inst{1}
          \and
          S.~Rehberg\inst{1}
          }
   \authorrunning{Avramenko et al.}

   \institute{University of Kiel, Institute of Theoretical Physics and Astrophysics, Leibnizstrasse 15, 24118 Kiel, Germany\\
              \email{[ravramenko;wolf;tillense]@astrophysik.uni-kiel.de}}

   \date{}

 
  \abstract
  { Inner cavities, accretion arms, and density waves are characteristic structures in the density distribution of circumbinary disks. 
   They are the result of the tidal interaction of the non-axisymmetric gravitational forces of
   the central binary with the surrounding disk and are most prominent in the inner region, where the asymmetry is most pronounced.}
   {The goal of this study is to test the feasibility of reconstructing the gas density distribution and quantifying properties of 
   structures in the inner regions of  edge-on circumbinary disks using multiple molecular line observations.}
   {The density distribution in circumbinary disks is calculated with 2D hydrodynamic simulations. Subsequently, 
   molecular line emission maps are generated with 3D radiative transfer simulations. Based on these, we investigate the observability 
   of characteristic circumbinary structures located in the innermost region for spatially resolved and unresolved disks.}
   {We find that it is possible to reconstruct the inner cavity, accretion arms, and density waves from spatially resolved 
   multi-wavelength molecular line observations of circumbinary disks seen edge-on. For the spatially unresolved observations only, an
   estimate can be derived for the density gradient in the transition area between the cavity and the disk's inner rim.}
   {}


   \maketitle
%

   \section{Introduction}
   
   Circumbinary disks make up a significant fraction of 
    protoplanetary disks \citep{Duquennoy_Mayor_1991, Kraus_2009}.  The additional torque provided by the binary is expected to alter  various physical
   processes in protoplanetary disks, such as planetesimal growth \citep{Marzari_2013} and planetary migration behavior  \citep{Kley_Haghighipour_2014}. 
   Inner regions are expected to show characteristic structures resulting from tidal interactions between the time-dependent non-axisymmetric 
   gravitational  potential generated by the binary and the surrounding disk. Typical structures are  cavities in the disk 
   center, accretion arms extending into these cavities, and density waves at the inner rim of the disk. All of these features are potentially most pronounced in the innermost regions 
   where the gravitational potential of the stellar binary shows the strongest deviation to that of a single star. However, not all of these structures are unique. 
   For example, inner cavities are also present in  transitional disks \citep{Garufi_2013, Dong_2017}, are caused by  photoevaporation 
   \citep{Clarke_2001}, or are due to the presence of a Jupiter-mass planet \citep{ Marel_Cazzoletti_2016, Marel_Dishoeck_2016}. In contrast to photoevaporation, the tidal force generated by a binary 
   is expected to affect the dust and gas phase equally. Moreover, due to the stronger tidal forces, the impact is more pronounced than in the case of a planet.

   In the case of a disk seen in face-on orientation, the structures resulting from binary-circumbinary disk interactions are expected to be detectable using proper combinations of infrared and 
   sub-millimeter wavelength continuum observations \citep{Avramenko_Wolf_2017, Ruge_2015}. 
   However, disks inclined edge-on obscure the innermost disk regions at optical to infrared wavelengths. Even at submm/mm wavelengths, the disk may still be optically thick in the continuum,
    as can be seen in the case of the Butterfly Star \citep{Grafe_2013}, where a flux reduction for submm wavelengths is observed in the disk's central region. 
    This reduction is attributed to the dust grain growth and vertical settling of larger dust grains.
   In contrast to continuum observations, molecular line emission has the advantage of the Doppler shift
   in frequency, which makes the optical thickness dependent on the velocity distribution of gas in the disk. By choosing proper molecular lines (i.e., isotopologs and transitions), it is therefore
   possible to directly observe the inner disk regions. Such observations were conducted by \cite{Dutrey_2017}, who detected a gap in the gas density of the "Flying Saucer" disk.
   A prominent example of an edge-on circumbinary disk cavity in the center is HH30. The cavity is supposed to be caused by a  binary with 0.31 $M_{\odot}$ and 0.14 $M_{\odot}$ and a separation of 
   18 au \citep{lynch_2019}. \cite{Louvet_2018} proposed an alternative explanation of the cavity based on the modeling of  dust continuum and molecular transition line ALMA 
   (Atacama Large Millimeter/submillimeter Array) observations. 
   According to \cite{Louvet_2018}, the cavity could be the result of a bow shock propagating outward.

   The goal of our study is to evaluate the feasibility of identifying characteristic binary-induced structures in edge-on circumbinary disks  using spatially resolved or unresolved molecular line observations.
   To address this goal, we conducted 2D hydrodynamic simulations of circumbinary disks with the code \texttt{Fosite} \citep{Illenseer_2009}. 
    The resulting density profiles were used to calculate the temperature of the system with the 3D grid-based Monte Carlo code 
   \texttt{Mol3D} \citep{Ober_2015}. This code was further applied to calculate the level populations for individual molecular lines and subsequently
   combined with the velocity fields to generate molecular line emission maps. Based on these maps, we evaluated the feasibility of detecting the cavity, accretion arms, and
   density wave.

   The paper is divided into two parts. In the first part (Sect.~\ref{sec:sim}), we give a short overview of the applied density distributions and radiative transfer
   simulations. Next, we discuss the simulated observations of the resulting maps in spatially resolved and unresolved cases (Sect.~\ref{sec:res}). The prime 
   focus lies on the possibility of reconstructing characteristic structures for the circumbinary disks.

   \section{Density distributions and radiative transfer simulations} 
   \label{sec:sim}

   In this section, we give a brief overview of the software used to conduct the hydrodynamic and radiative transfer simulations.
   A complete discussion can be found in \citep{Avramenko_Wolf_2017}.
   \subsection{Density distributions}

   The density and gas velocity distributions were calculated using the 2D grid-based hydrodynamic code   \texttt{Fosite} \citep{Illenseer_2009}. A full discussion of the simulation setup, 
   as well as the parameter space, is available in \cite{Avramenko_Wolf_2017}, where we investigated the feasibility of observing 
   binary-induced disk structures in the continuum. 
   For the current study, we  used a single hydrodynamic simulation of this sample,
   namely a disk with a semimajor axis of $a = 30$ au and 
   binary masses of $M_{\rm prim} = M_{\rm sec} = 1{\rm M_{\odot}}$, with $M_{\rm prim}$ and $M_{\rm sec}$ as the masses of the primary and secondary components, respectively.
   As was found by  \cite{Avramenko_Wolf_2017}, these parameters result in particularly large 
   accretion arms and  density oscillations. Both factors increase the chances of detecting these features.

   Because  \texttt{Fosite} 
   only calculates the 2D surface density distribution, we reconstructed the vertical density profile of the disk using the geometrical thin disk model \citep{LyndenBell_1969, Shakura_1973}. 
   The scale height of the disk was calculated during the hydrodynamic simulation under the assumption of a vertically isothermal disk model. 
   To get the appropriate scale height, we iterated between (a) calculating the scale height with \texttt{Fosite} using a given temperature distribution and (b) subsequently calculating the
   temperature with \texttt{Mol3D} \citep{Ober_2015} for the derived vertical disk structure. This iteration was repeated until the change in the resulting scale height 
   between two subsequent iterations was below 1 percent. The initial temperature distribution was set to 50 K (isothermal).

   The  vertical  component of the velocity cannot be deduced from the simulations, and we assume that it is small 
   compared to the orbital velocity (i.e., primarily governed by Keplerian motion). An indirect contribution 
   was made, however, by including a general line broadening due to turbulence in the disk. We will further discuss this effect in the next section.
  The full parameter set is compiled in Table~\ref{fig:hyd_param}.

    \begin{table*}
  \caption{Parameters of the hydrodynamic simulation.}
\begin{tabular}{lc|c}
        \multicolumn{2}{c}{}                                                                                \\ \hline \hline
       \bf{Parameter}                       &                                      &  \bf{Parameter value}  \\ \hline
       Mass of individual binary components &   $M_{\rm prim} = M_{\rm sec}$ [${\rm M_{\odot}}$]    & $1$   \\ \hline
       Initial disk mass                    &   $M_{\rm disk}$ [${\rm M_{\odot}}$] & $0.02 $                \\ \hline
       Semi-major axis                      &   $a$ [au]                           & $30$          \\ \hline
       Inner boundary radius                &   $r_{\rm in}$ [au]                  & $0.6 \cdot a$ \\ \hline
       Outer boundary radius                &   $r_{\rm out}$ [au]                 & $1000$        \\ \hline
       Eccentricity                         &   $\varepsilon$                      & $0$                    \\ \hline
       Simulation time                      &   $t_{\rm sim}$ [yr]                 & $5 \cdot 10^{5}$       \\ \hline \hline
 \end{tabular}
 
  \label{fig:hyd_param}
  \end{table*}
    
    In preparation of the analysis of the circumbinary disk model, we also considered an analytically described density distribution of a Keplerian circumstellar disk (see Eq.~\ref{eq:1}),
    with the quantity $\rho$ being the volume density, $\Sigma$ the surface density, $H$ the pressure scale height, and $z$ the vertical disk coordinate. 
    The central star in the Keplerian case has the combined mass of the binary ($2{\rm M_{\odot}}$) of the circumbinary case.
    First studying the Keplerian disk, as well as selected modified versions (e.g., the inner cavity and gap),
   will help us to identify distinct features that are associated with a specific morphology of the disk:
   
       \begin{equation} \label{eq:1}
  \varrho = \frac{\Sigma}{\sqrt{2 \pi}H} \, \exp\Big[-\frac{1}{2}\left(\frac{z}{H}\right)^2\Big]  
  .\end{equation}

   \subsection{Radiative transfer simulation}

   The density distribution calculated with \texttt{Fosite} was further used to calculate the corresponding temperature distribution. 
   This was done with the Monte Carlo-based 3D continuum and line radiative transfer code \texttt{Mol3D} \citep{Ober_2015}. 
   As above, we refer to \cite{Avramenko_Wolf_2017} for a full discussion of the radiative transfer simulations.
   The parameters used in the current study are summarized in Table~\ref{fig:rad_param}.
   
    \begin{table*}
   \caption{Parameters of the radiative transfer simulation.}
\begin{tabular}{lc|c}
        \multicolumn{2}{c}{}                                                                                                                   \\ \hline \hline
         \bf{Parameter}                 &                                                              & \bf{Parameter value}                  \\ \hline
         Stellar effective temperature  & $T_{ \rm prim} = T_{ \rm sec}$ [K]                           & $ 5600 $                              \\ \hline
         Stellar radius                 & $R_{ \rm prim} = R_{ \rm sec}$ [$R_{\odot}$]                 & $ 0.98$                               \\ \hline
         Stellar luminosity             & $L_{ \rm prim} = L_{ \rm sec}$ [$M_{\odot}$]                 & $ 0.87$                               \\ \hline
         Inner disk radius              & $r_{\rm in}$ [au]                                   & $ 0.6 \cdot a = 0.6 \cdot 30 = 18 $   \\ \hline
         Outer disk radius              & $r_{\rm out}$ [au]                                  & $ 200$                                \\ \hline
         Simulation wavelengths         & $\lambda_{\rm sim}$ [$\rm \mu m$]                            & $ 0.05 \dots 2000$                    \\ \hline \hline
 \end{tabular}

  \label{fig:rad_param}
  \end{table*}

   The synthetic molecular line maps for the individual velocity channels were calculated using the large velocity gradient (LVG) method \citep{Pavlyuchenkov_2007}. 
    For the line broadening, we assumed a model based on earlier line observations by, for example, 
    \cite{Pietu_2007} and \cite{Chapillon_2012} according to the formula 
   
   \begin{equation} \label{eq:2}
    \bigtriangleup V = \sqrt{ \frac{2kT}{m} + v_{\textrm{turb}}^2} \, ,
    \end{equation}
   
   \noindent
   with $\bigtriangleup V$ being the total line broadening, $k$ the Boltzmann constant, $T$ the temperature, $m$ the molecular mass, and $v_{\textrm{turb}}$ the turbulent contribution. 
   \cite{Pietu_2007} assumed $v_{\textrm{turb}}$ to be a power law with regard to the radius. To reduce the number of free parameters, we instead used a constant value of $100$ m/s.

\begin{figure}[h!]
         \resizebox{\hsize}{!}{\includegraphics{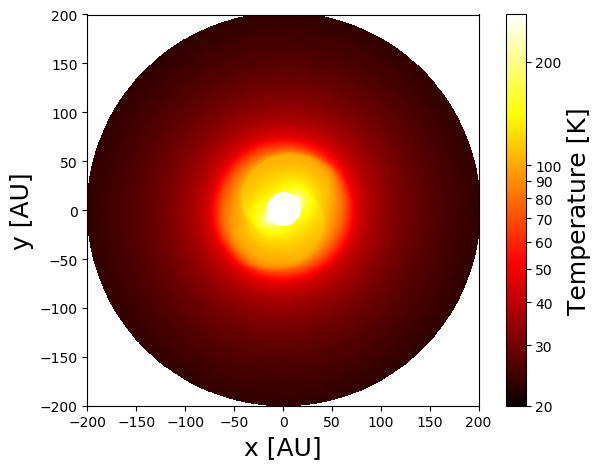}}
         \caption{Midplane temperature distribution for a circumbinary disk.}
         \label{fig:temp_mid_plane}
    \end{figure}

    For our study, we used two different isotopologs of the carbon monoxide molecule: ${\rm C}{\rm O}$ and ${\rm C}^{18}{\rm O}$.
     For the abundances we assumed molecule-to-hydrogen ratios (N/N(H)) of $10^{-4}$ for ${\rm C}{\rm O}$ and $2 \cdot 10^{-7}$ for ${\rm C}^{18}{\rm O,  }$ as reported in \citep{Pavlyuchenkov_2007}.
    These were calculated under the assumption of the gas-grain chemical model with surface reactions from \cite{Semenov_2004, Semenov_2005}. The temperature in the disk midplane is above $20$ K
     (Fig.~\ref{fig:temp_mid_plane}), which should prevent CO from freezing out.

   Transition frequencies and transition rates for the considered isotopologs are taken from the Leiden Atomic and Molecular Database (LAMDA; \citealt{Schoier_2005}).
   The first three rotational transition lines for each of these isotopologs, $J = 2-1$, $J = 3-2$, and $J = 4-3$,  were considered. 
   We used a maximum velocity of 6.5 ${\rm km/s}$ for the Keplerian circumstellar disk and 7.0 ${\rm km/s}$ for the circumbinary disk, with 67 and 99 velocity channels, respectively.
   This results in a spectral resolution that is comparable to the maximum resolution achievable with ALMA in the corresponding bands.

   \section{Results}
   \label{sec:res}

   This section is divided in two parts. First we analyze   simulated "spatially resolved" 
   molecular line maps. We begin with  a standard Keplerian disk as a reference case of  a featureless disk. After that we discuss the case of a Keplerian 
   disk with a large inner cavity and another with a gap in the density 
   distribution located between 40 and 50 au. We proceed with a disk showing a  radial wave structure in its density. Finally, we discuss the case of a circumbinary disk.

   The second part deals with "spatially unresolved" observations. There we follow the same sequence in the discussion of our results (i.e., starting with the circumstellar case). Afterward, we connect the 
   distinctive elements in our data with the density gradient expected to occur at the inner edge of a circumbinary disk.
   
   The midplane of the disk coincides with the x-y plane. We are discussing edge-on disks with 
   the observer's line of sight along the y axis. All models are presumed to be located at a distance of 140 pc.
   
   \subsection{Spatially resolved line observations}
   
   \subsubsection{Keplerian disk}
   \label{subsec:res_resolved_kep}

\begin{figure*}[ht]
  
         \begin{subfigure}{.5\textwidth}
             \includegraphics[width=1\linewidth]{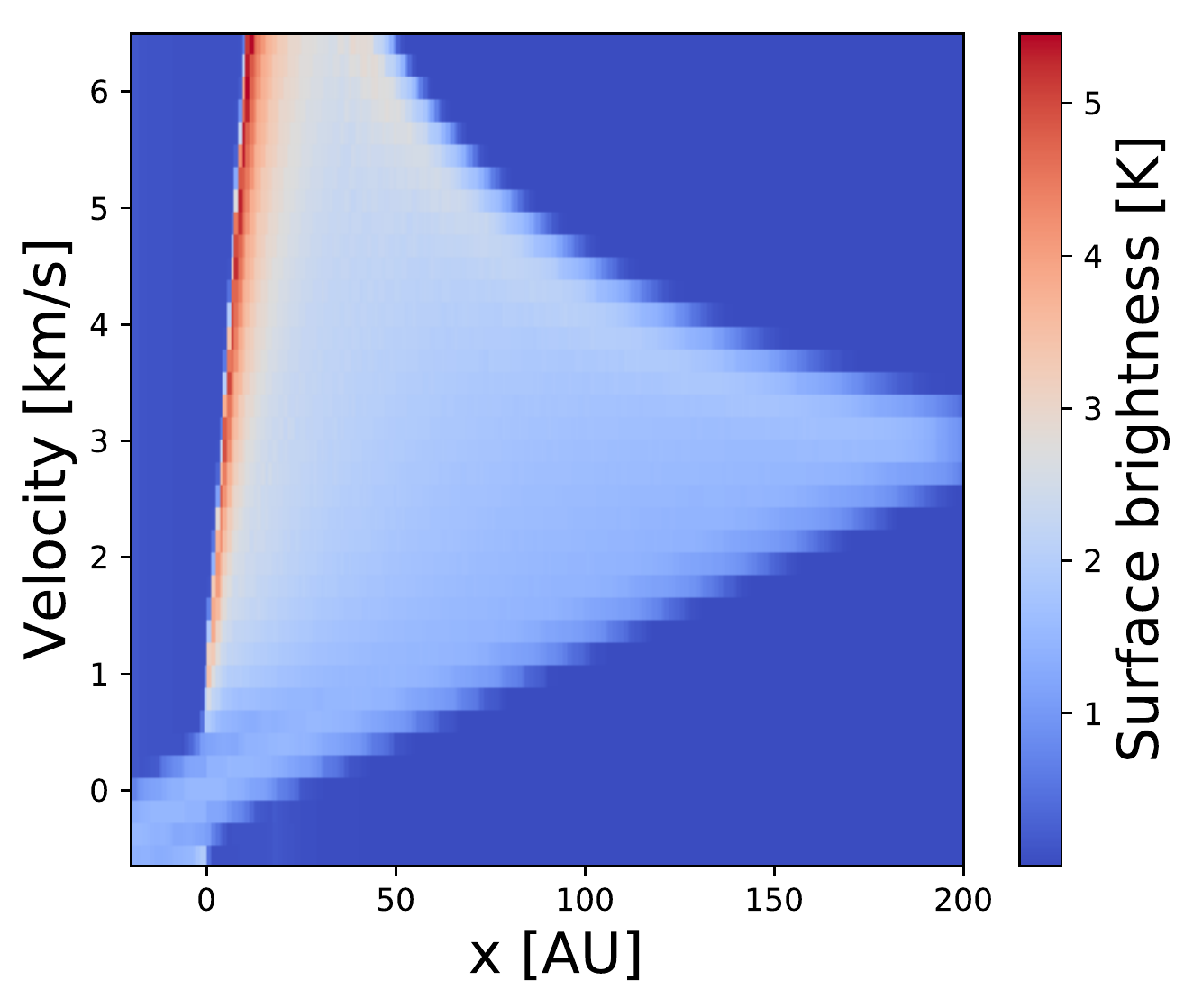}
              \subcaption{}
              \label{fig:kepler_line_30_co_1}
         \end{subfigure}
         \begin{subfigure}{.5\textwidth}
             \includegraphics[width=1\linewidth]{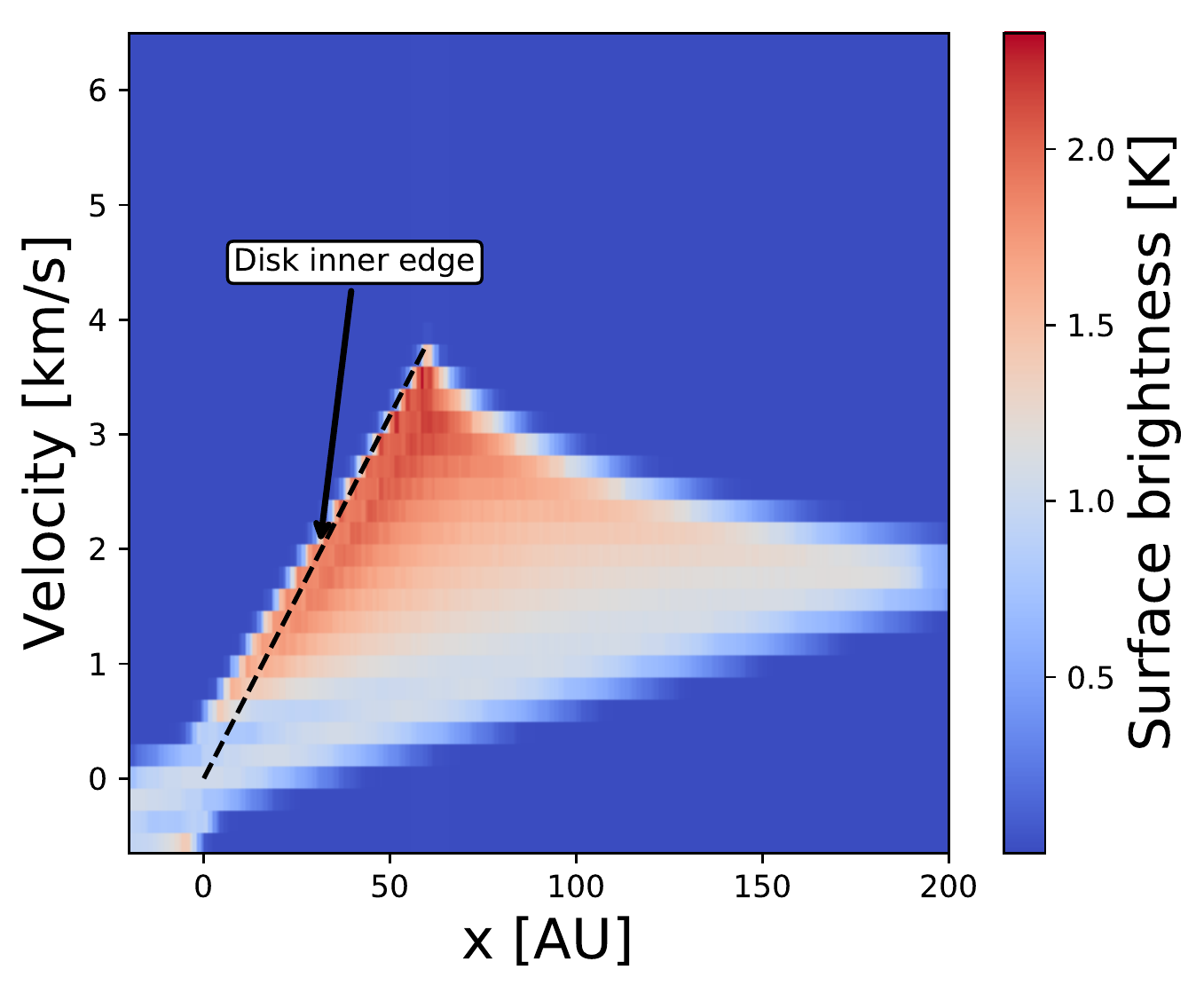}
             \subcaption{}
             \label{fig:kepler_cavity_line_30_co_1}
         \end{subfigure}
         \begin{subfigure}{.5\textwidth}
             \includegraphics[width=1\linewidth]{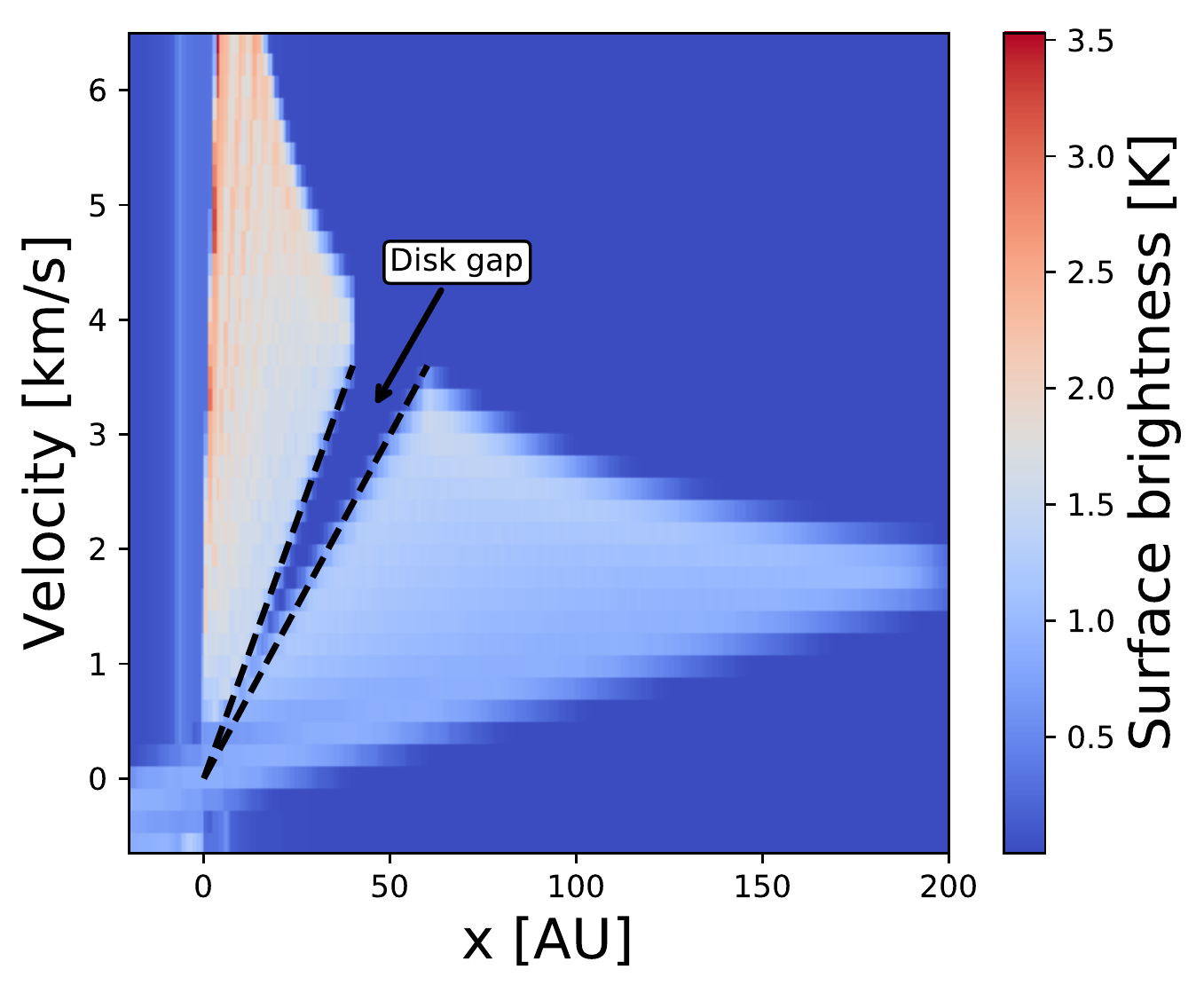}
             \subcaption{}
             \label{fig:kepler_gap_line_30_co_1}
         \end{subfigure}
         \begin{subfigure}{.5\textwidth}
             \includegraphics[width=1\linewidth]{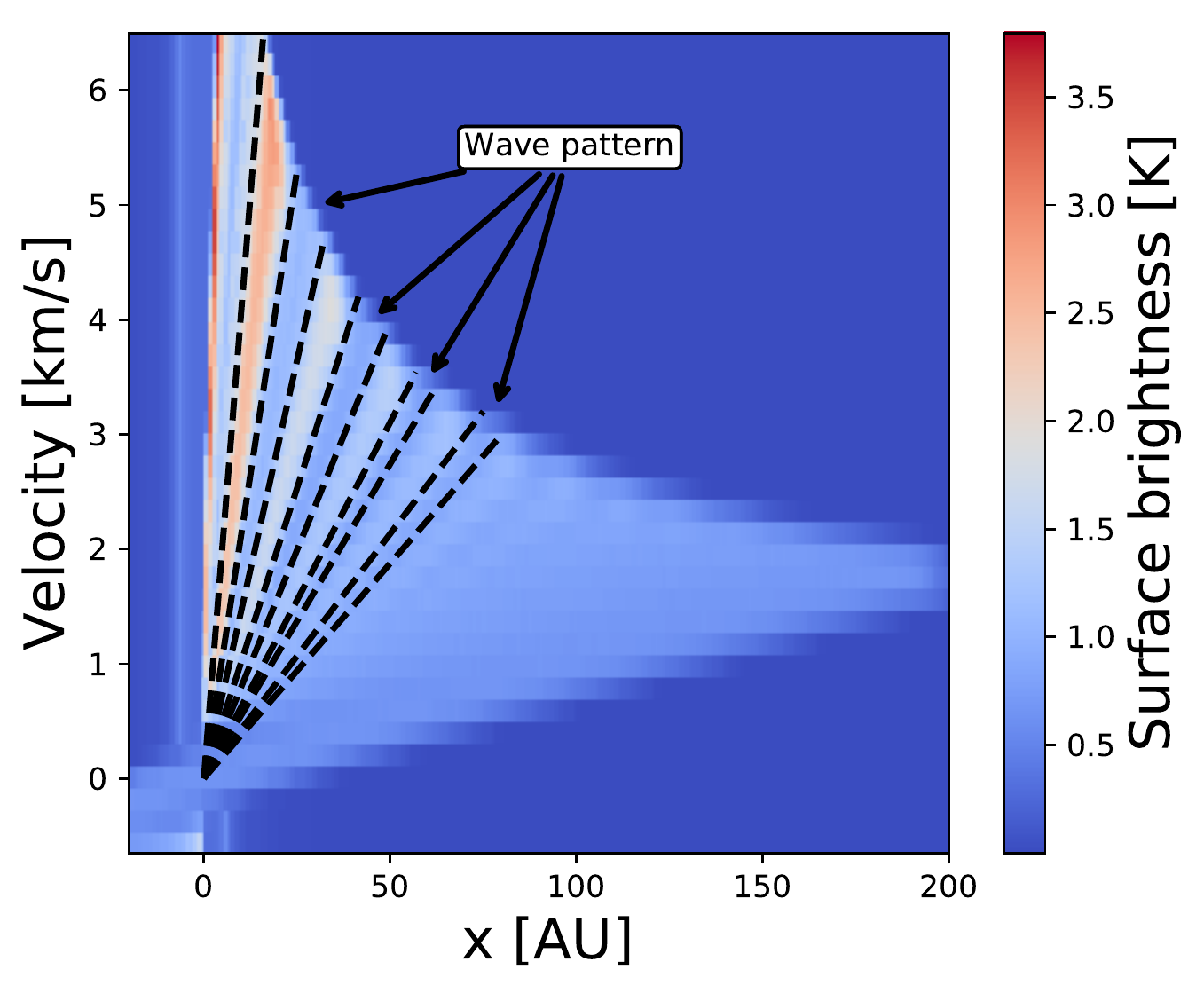}
             \subcaption{}
             \label{fig:kepler_den_wave_line_30_co_1}
         \end{subfigure}
            \caption{ ${\rm C}{\rm O}$  $J = 4-3$  Velocity-resolved surface brightness distribution of a Keplerian disk seen edge-on
            (see Sect.~\ref{subsec:res_resolved_kep} for details).
            (a) No features.
            (b) Cavity at $r_{\rm in} = 60$ au. The inclined black dashed line indicates the lack of flux inside the cavity.
            (c) Gap between $r_{\rm in} = 40$ and $60$ au. The inclined black dashed lines indicate the lack of flux inside the gap.
            (d) $Sin$-wave density profile. Black lines indicate the flux maxima (minima) corresponding to the location of maxima (minima) of the density distribution.
          }
            \label{fig:kepler_line}
      \end{figure*}

  Our simulation results are stored in a 3D data cube with two spatial dimensions (along the x and z axes) for each of the velocity channels. In this study, we are interested in the spatial 
  distribution of gas in the disk. To achieve better contrast and reduce the number of necessary plots, we integrated the flux density along the vertical extent of the edge-on viewed disk 
  and divide it by the number of contributing pixels  
  (see Fig.~\ref{fig:kepler_line_30_co_1} for an example of  an unmodified edge-on Keplerian disk). 
  Due to Keplerian motion, each part of the disk moves with a specific projected velocity to the line of sight 
  of the observer, resulting in a frequency shift. The projected velocity depends on the position along the x axis. 
  The values of the flux along a straight line through the origin correspond to the radiation of the gas at a specific radius in the disk. 
  To determine the radius, one notes the x value in a velocity channel at which the flux 
  falls to zero (for the derivation, see Appendix A in \citealt{Dutrey_2017}). The contributions from different radii are given along lines with different slopes  in the diagram.

      This interpretation of the velocity-position diagram is illustrated by the case of  a Keplerian disk with a large cavity 
      from which the disk's inner radius of 60 au and the corresponding Keplerian velocity of $\sim$ 3.6 ${\rm km/s}$ can be derived (Fig.~\ref{fig:kepler_cavity_line_30_co_1}).
      As both sides of the edge-on disk are observed,  the extent of the cavity in each direction can be determined separately (important for 
      disks with non-circular cavities).

   In the case of the Keplerian disk with a radial gap, the inner and outer radii of this gap can be identified  (see Fig.~\ref{fig:kepler_gap_line_30_co_1}).
   Here, the location and extent of the gap can be directly deduced from the velocity-position diagram as well.
   A similar observation was already made by \cite{Dutrey_2017}, resulting in the identification of a gap in the Flying Saucer  disk 
   using ${\rm C}{\rm O}$ $J = 2-1$  observations.

   In the final example, the Keplerian disk density distribution  is modified with a $sin$-function to mimic a density wave propagating radially outward. 
   The resulting velocity-position diagram (Fig.~\ref{fig:kepler_den_wave_line_30_co_1}) is similar to that generated by a superposition of "gap-like" features with varying radii.
   In this case, it would be possible to determine the wavelength of the density oscillation by measuring the distance between the flux minima (or maxima) in a similar manner, as  
   in the case of the analysis of a gap structure.

   \subsubsection{Circumbinary disk}
   \label{subsec:res_resolved_bin}

\begin{figure*}[ht]
  
         \begin{subfigure}{.5\textwidth}
             \includegraphics[width=1\linewidth]{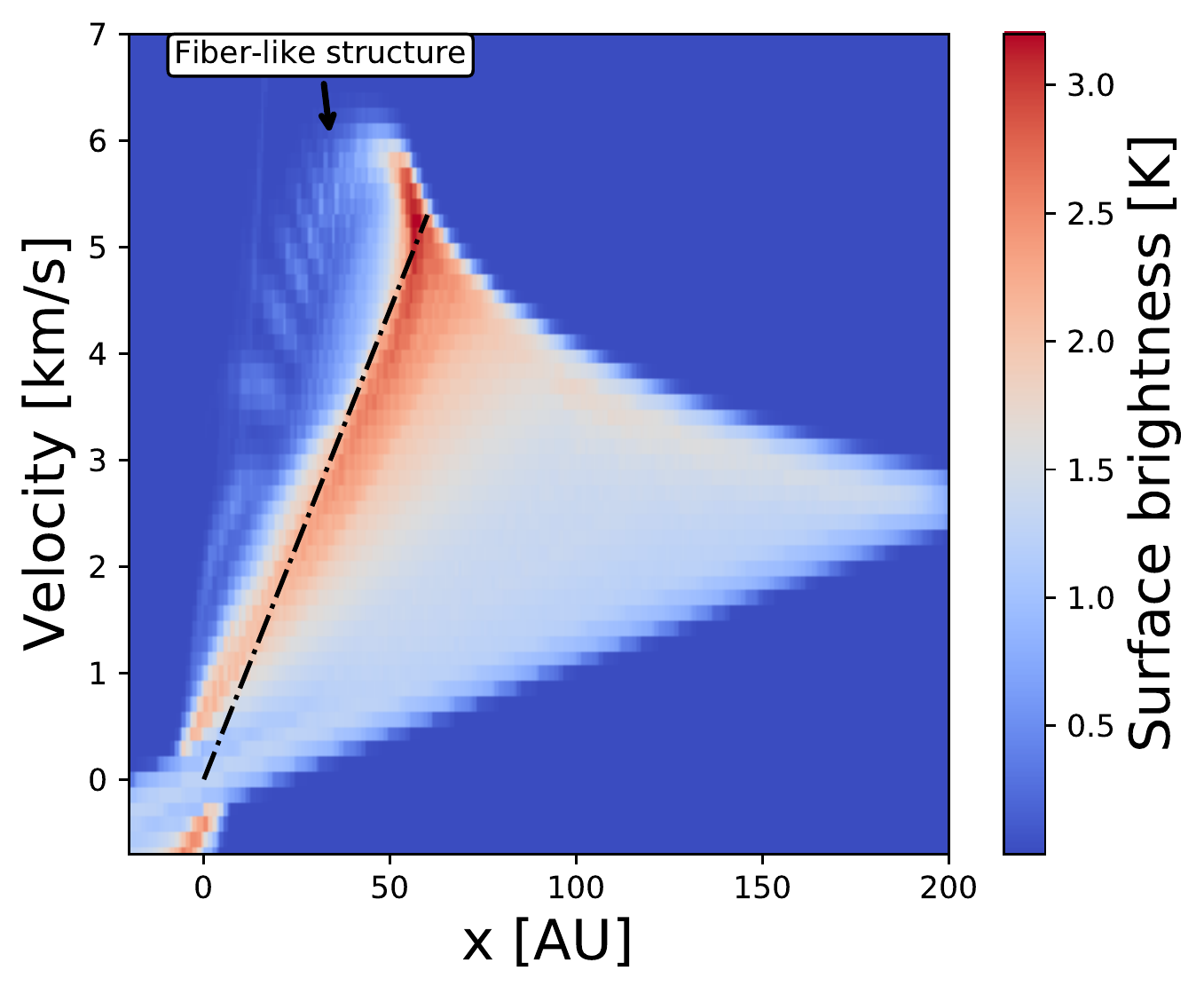}
              \subcaption{}
              \label{fig:y-int_co_1}
         \end{subfigure}
         \begin{subfigure}{.5\textwidth}
             \includegraphics[width=1\linewidth]{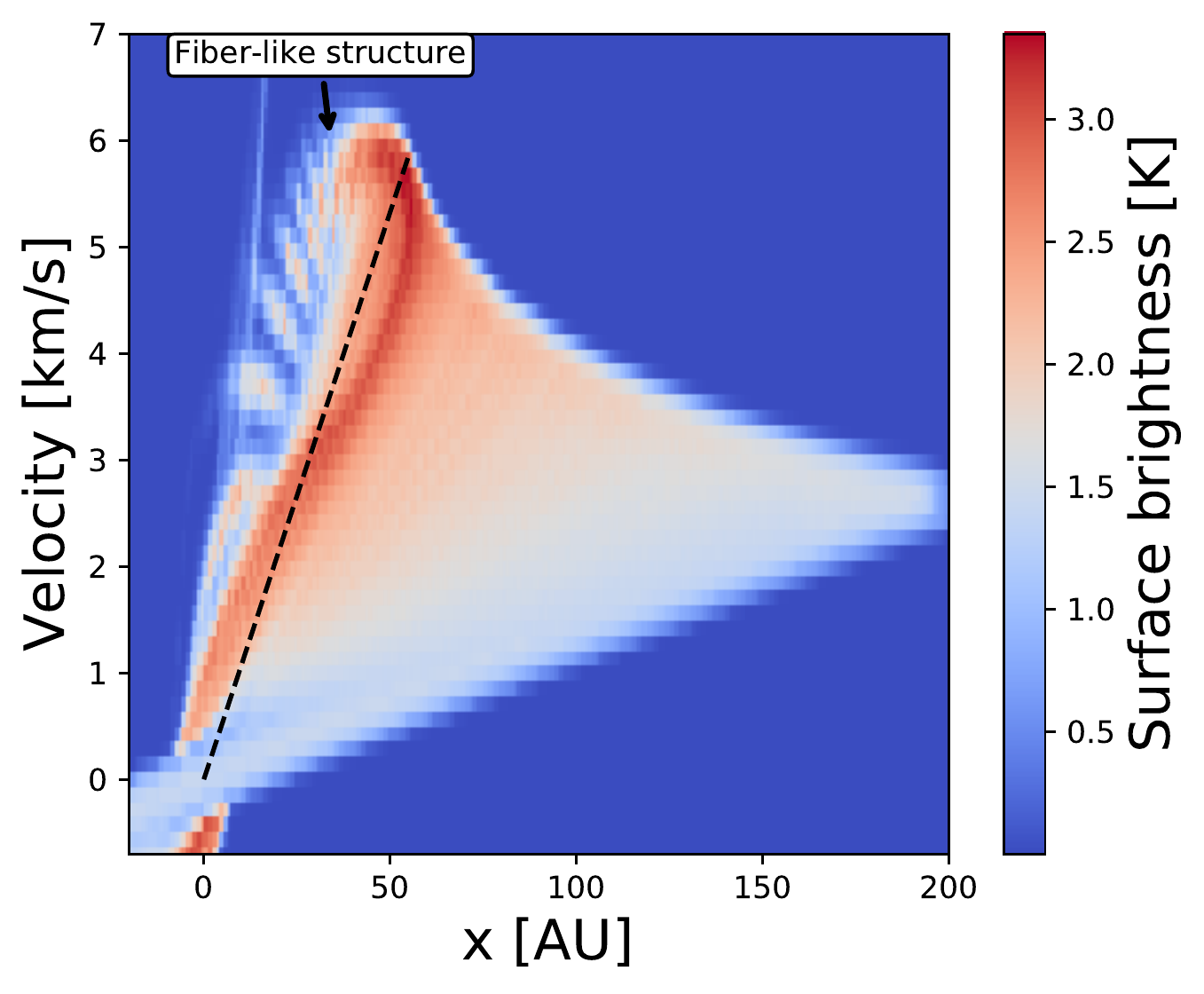}
             \subcaption{}
             \label{fig:y-int_co_3}
         \end{subfigure}
         \begin{subfigure}{.5\textwidth}
             \includegraphics[width=1\linewidth]{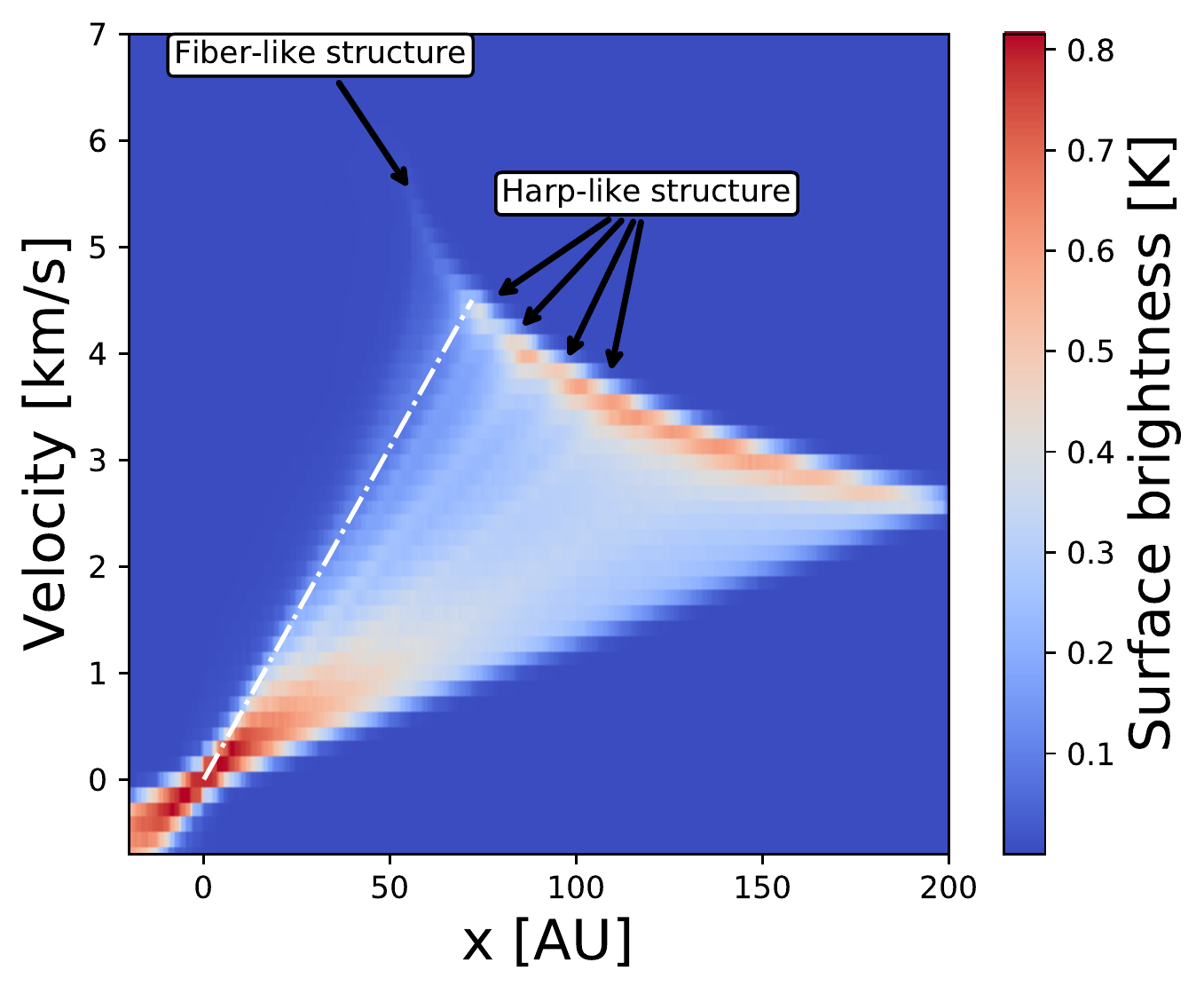}
             \subcaption{}
             \label{fig:y-int_c18o_1}
         \end{subfigure}
         \begin{subfigure}{.5\textwidth}
             \includegraphics[width=1\linewidth]{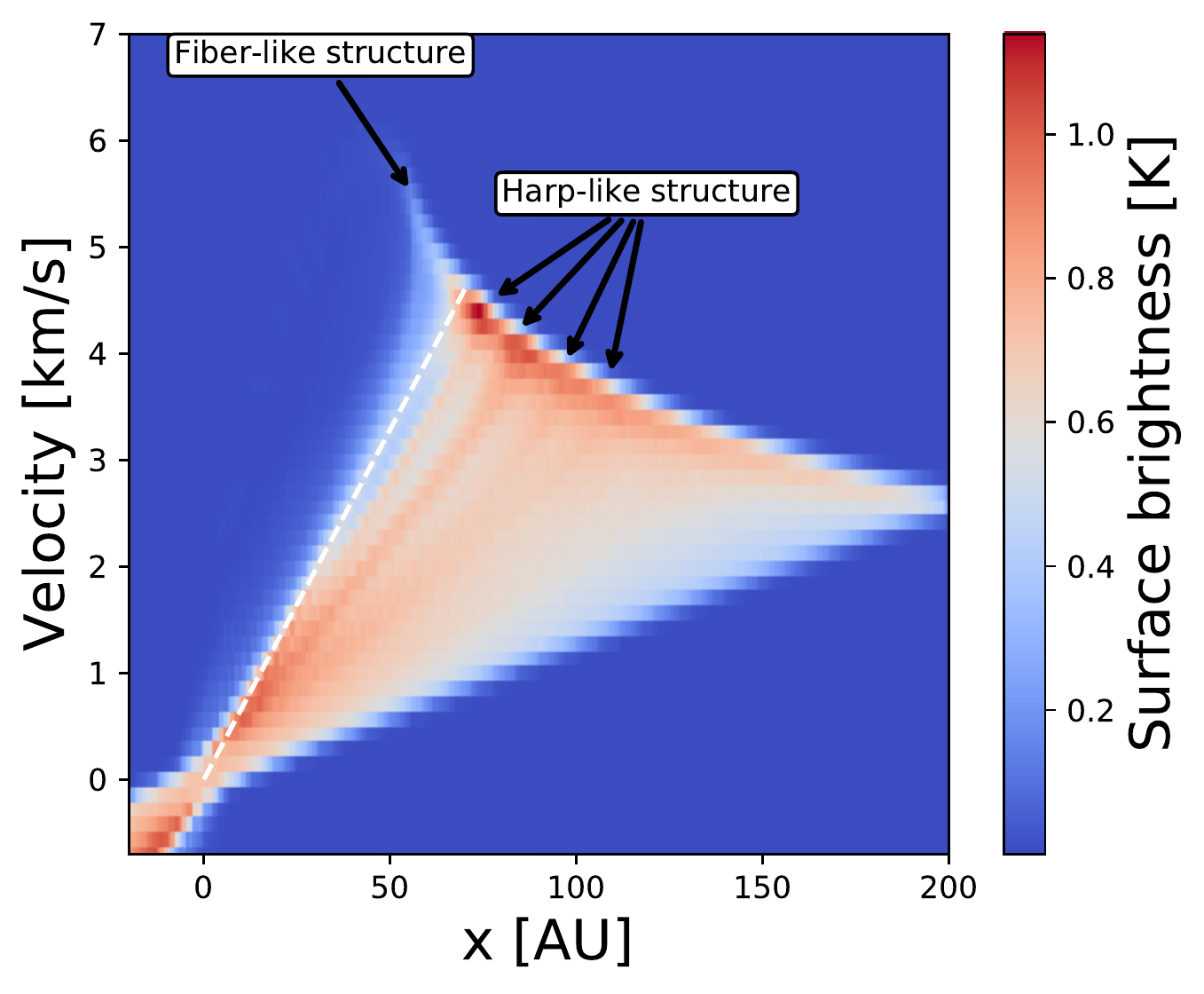}
             \subcaption{}
             \label{fig:y-int_c18o_3}
         \end{subfigure}
            \caption{   Velocity-resolved surface brightness distribution of a circumbinary disk (see Sect.~\ref{subsec:res_resolved_bin} for details).
            (a) ${\rm C}{\rm O}$ $J = 2-1$. \\
            (b) ${\rm C}{\rm O}$ $J = 4-3$.
            (c) ${\rm C}^{18}{\rm O}$ $J = 2-1$.
            (d) ${\rm C}^{18}{\rm O}$ $J = 4-3$.
            }
            \label{fig:binary_line}
      \end{figure*}

\begin{figure}[h!]
          \centering
           \includegraphics[width=0.95\columnwidth]{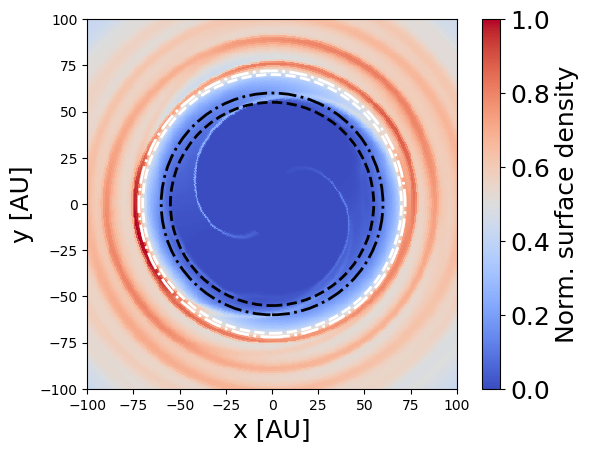}                       
          \caption{Derived inner radii plotted onto the initial surface density distribution.
          ${\rm C}{\rm O}$      - $J = 2-1$ (black, dash-dotted);
          ${\rm C}{\rm O}$      - $J = 4-3$ (black, dashed);
          ${\rm C}^{18}{\rm O}$ - $J = 2-1$ (white, dotted);
          ${\rm C}^{18}{\rm O}$ - $J = 4-3$ (white, dashed).
          }
          \label{fig:density_circle}
       \end{figure}

          Now we focus our attention on the case of a circumbinary disk. Fig.~\ref{fig:binary_line} show the vertically integrated fluxes 
           for transition lines $J = 2-1$ 
          and  $J = 4-3$ of ${\rm C}{\rm O}$ and ${\rm C}^{18}{\rm O}$. As expected, we observe low fluxes in the region close to the central star (i.e., small radii) and thus high velocities.
          This is a clear indication of the cavity associated with the circumbinary disk. However, 
          as can be best seen in the case of the ${\rm C}{\rm O}$ $J = 4-3$ line, the cavity is not completely empty. One can clearly see the fiber-like structure 
          that can be attributed to the presence of accretion arms in the disk.  In fact, if we extract the inner radius of the disk from the fluxes of each CO isotopolog and each line, we derive the inner radii shown in 
          Fig.~\ref{fig:density_circle}. Here we can clearly see that the more abundant isotopolog ${\rm C}{\rm O}$ 
          leads to a smaller derived radius. We determine a radius of 60 and 50 au for $J = 2-1$ and $J = 4-3$, respectively.
          For ${\rm C}^{18}{\rm O}$, we derive inner  radii of 72 and 70 au for transition lines $J = 2-1$ and $J = 4-3$, respectively.

        The  isotopolog ${\rm C}^{18}{\rm O}$ also allows us to constrain the inner disk structure (see Figs.~\ref{fig:y-int_c18o_1} and~\ref{fig:y-int_c18o_3}). It shows a
        harp-like structure, which is caused by the density waves propagating through the disk, similar to the case discussed in 
        Sect.~\ref{subsec:res_resolved_kep} and Fig.~\ref{fig:kepler_den_wave_line_30_co_1}.
        If the spectral resolution is high enough,
        this allows one to detect the flux differences between the corresponding density  maxima and minima. 
          Moreover, since we observe multiple maxima and minima, it is possible to derive the wavelength of the oscillation. 
          As was shown by \cite{Avramenko_Wolf_2017}, the wavelength of the density wave is connected to the basic parameters of the binary system, such as binary mass  and binary separation.

   \subsection{Spatially unresolved line observations}
   \label{subsec:res_unresolved}

   In the previous section, we discussed the potential of spatially resolved line observations to constrain selected features of the disk morphology, such as the presence and shape of a central cavity.
  However, the angular resolution and/or brightness sensitivity of the observations may be insufficient to study this morphology. 
   For this reason,  we investigate the feasibility of deriving selected disk parameters using  spatially unresolved observations. To achieve this, we calculated the ratios between two spatially integrated molecular 
  lines. In one case, we combined data from the same isotopolog but for two different transition lines. In the second case, we used the same transition line $J$ but considered different isotopologs. This allows one
   to trace regions where the ratio shows a high degree of variation, which is expected to occur in areas with high temperature and density gradients inside the cavity. We will first 
   briefly discuss the case of a molecular line in an edge-on disk. After that we will address the line ratios for the simple case of a Keplerian disk. 
   This will allow us to better identify the features associated with the presence of a gap in the disk, followed by the case of a circumbinary disk.

\begin{figure}[h!]
          \resizebox{\hsize}{!}{\includegraphics{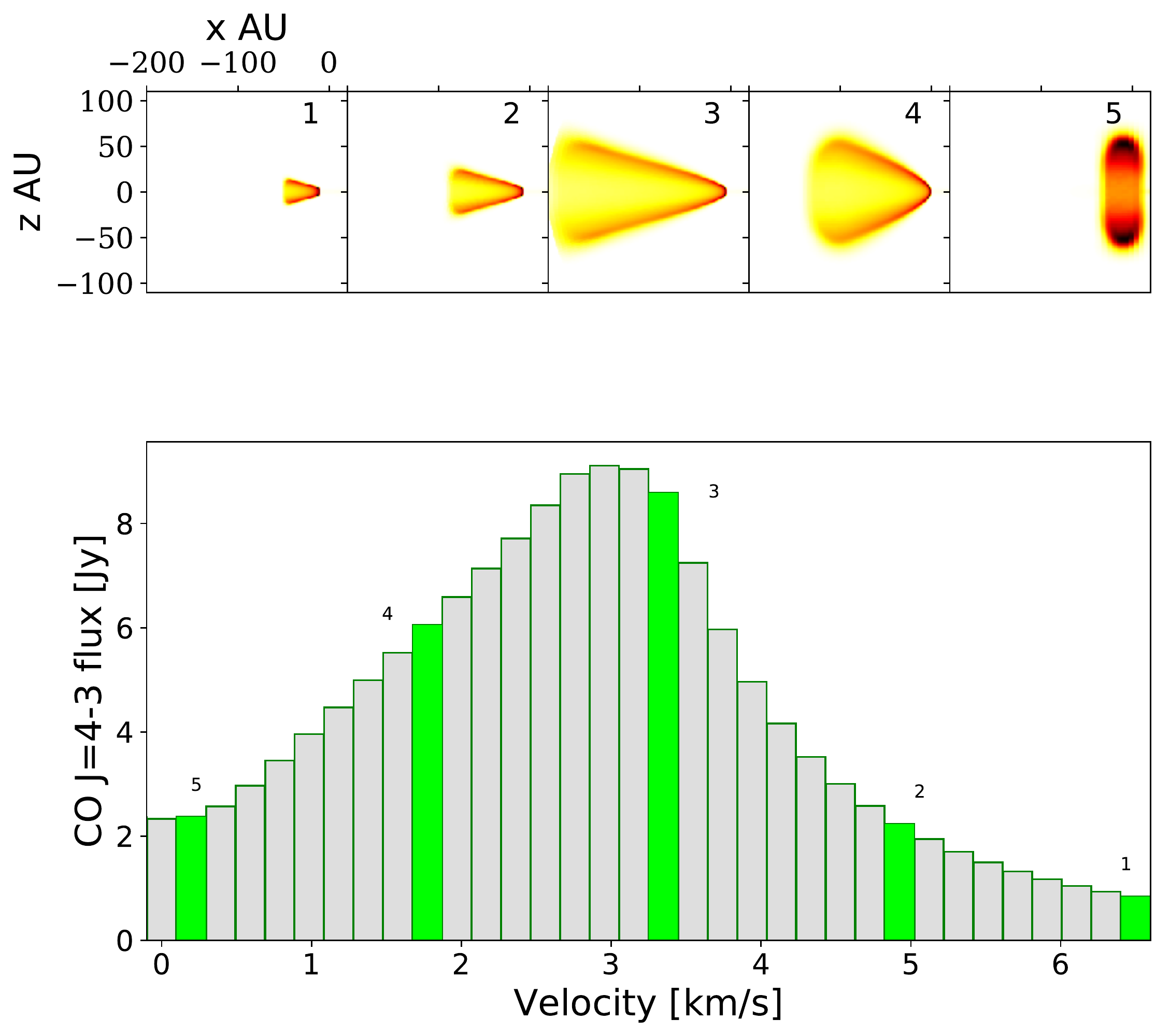}}
          \caption{ Keplerian disk line  ${\rm C}{\rm O}$,  $J = 4-3$ emission. \textit{Top:}  Spatially resolved  images for specific channels numbered  1 through 5. 
          \textit{Bottom:} ${\rm C}{\rm O}$ line profile. The channels from the top image are highlighted in blue and denoted by a number.}
          \label{fig:velocity_channel}
        \end{figure}

   In order to understand the shape of a molecular line, one has to understand the shape of the area of the disk in each velocity channel.
   A more detailed discussion of line formation in protoplanetary disks is available in Sect. 2.2 of \cite{Beckwith_1993}. The exact shape of the 
   covered area can be calculated with
   
      \begin{equation} \label{eq:2}
    r(\phi) =  \frac{GM_{\textrm{star}}}{v_{\textrm{obs}}} \sin^2{\theta}\cos^2{\phi}  \, ,
    \end{equation}

   \noindent
    with $G$ being the gravitational constant, $M_{\textrm{star}}$ the mass of the central object, $v_{\textrm{obs}}$ the average projected velocity toward the observer, 
   and $\phi$ and $\theta$ the angles in polar coordinates.
   Figure~\ref{fig:velocity_channel}  shows the generic CO $J=4-3$ line of a 
   Keplerian  disk seen edge-on. Additionally, the spatially resolved images of the edge-on disk are displayed for the highlighted velocity channels. The line 
   profile shows two distinct parts. The first part corresponds to the area covered by a singular velocity channel that
   moves outward in the disk (Numbers 1, 2, and 3 in Fig.~\ref{fig:velocity_channel}). 
   After reaching the peak flux, the area contributing to the flux starts moving toward the center of the disk (Numbers 4 and 5 in Fig.~\ref{fig:velocity_channel}). This behavior can be explained by the projection 
   of the Keplerian velocity on the line 
   of sight of the observer. The process repeats itself on the other side of the disk. In Fig. 1 of \cite{Beckwith_1993}, one can see the exact shape of the area covered by a velocity channel 
   in connection with the location at the line profile.  For the Keplerian disk, both sides are symmetric. This results in the line profiles as well as the line ratios being symmetric
   as well. Later we will see that the two sides of the line spectra are not exactly symmetric in the case of a binary disk. For that reason, we restrict figures of Keplerian disks to one side 
   of the spectrum. For binary cases, both sides are shown.
   
    \subsubsection{Keplerian disk}
    \label{subsec:res_unresolved_kep}

\begin{figure*}[ht]
  
         \begin{subfigure}{.5\textwidth}
             \includegraphics[width=1\linewidth]{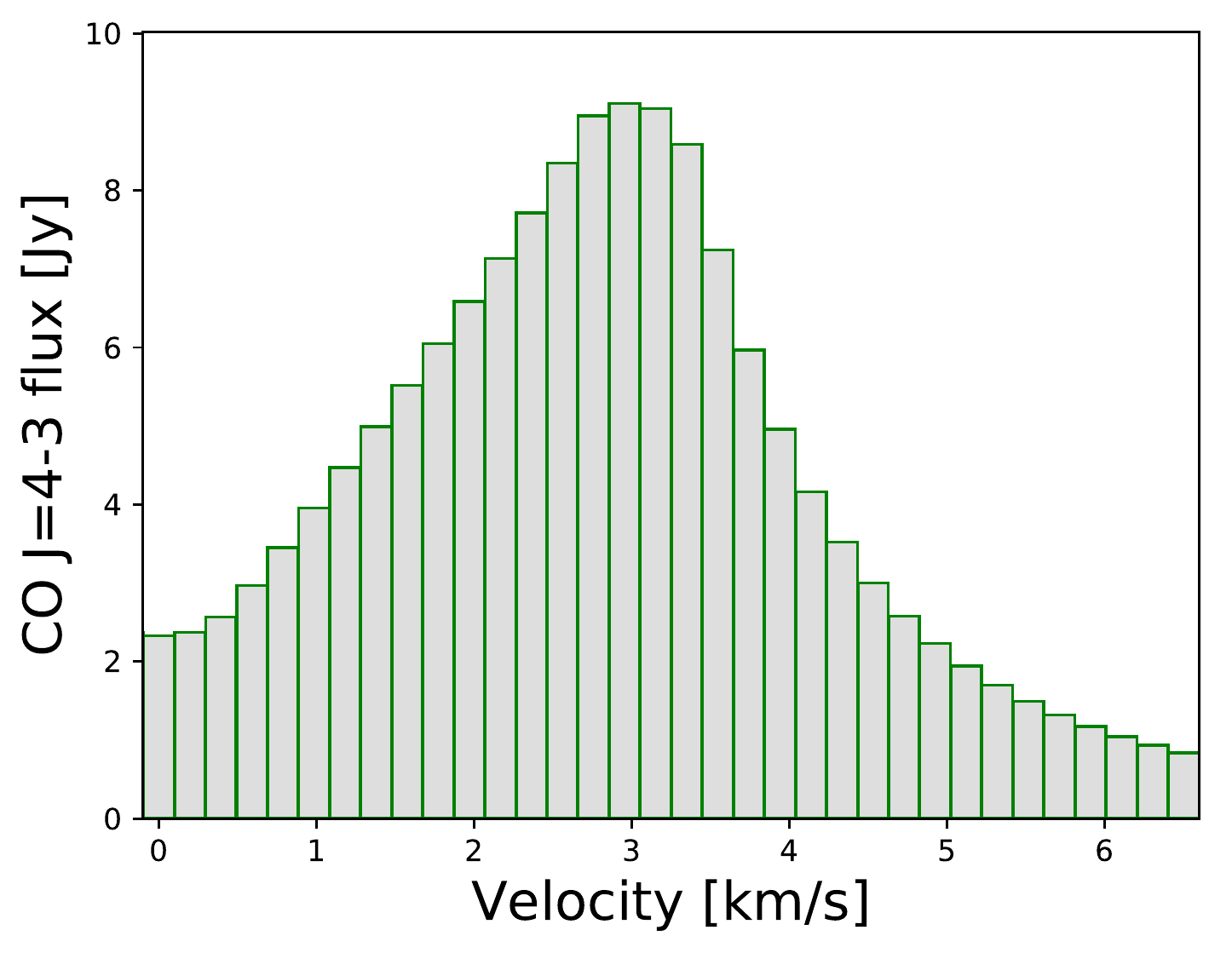}
              \subcaption{}
              \label{fig:kepler_co_line_4-3}
         \end{subfigure}
         \begin{subfigure}{.5\textwidth}
             \includegraphics[width=1\linewidth]{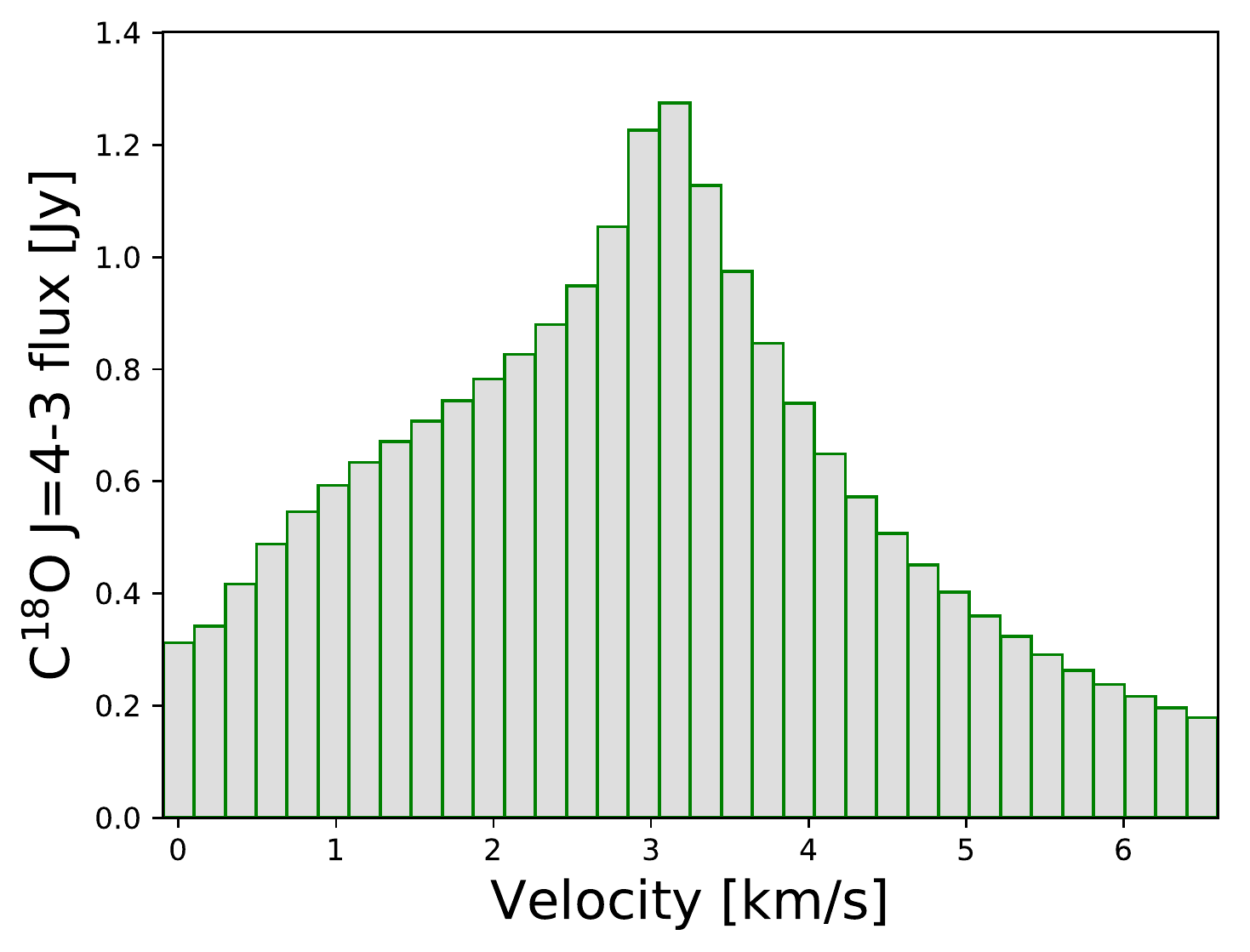}
             \subcaption{}
             \label{fig:kepler_c18o_line_4-3}
         \end{subfigure}
~
         \begin{subfigure}{.5\textwidth}
             \includegraphics[width=1\linewidth]{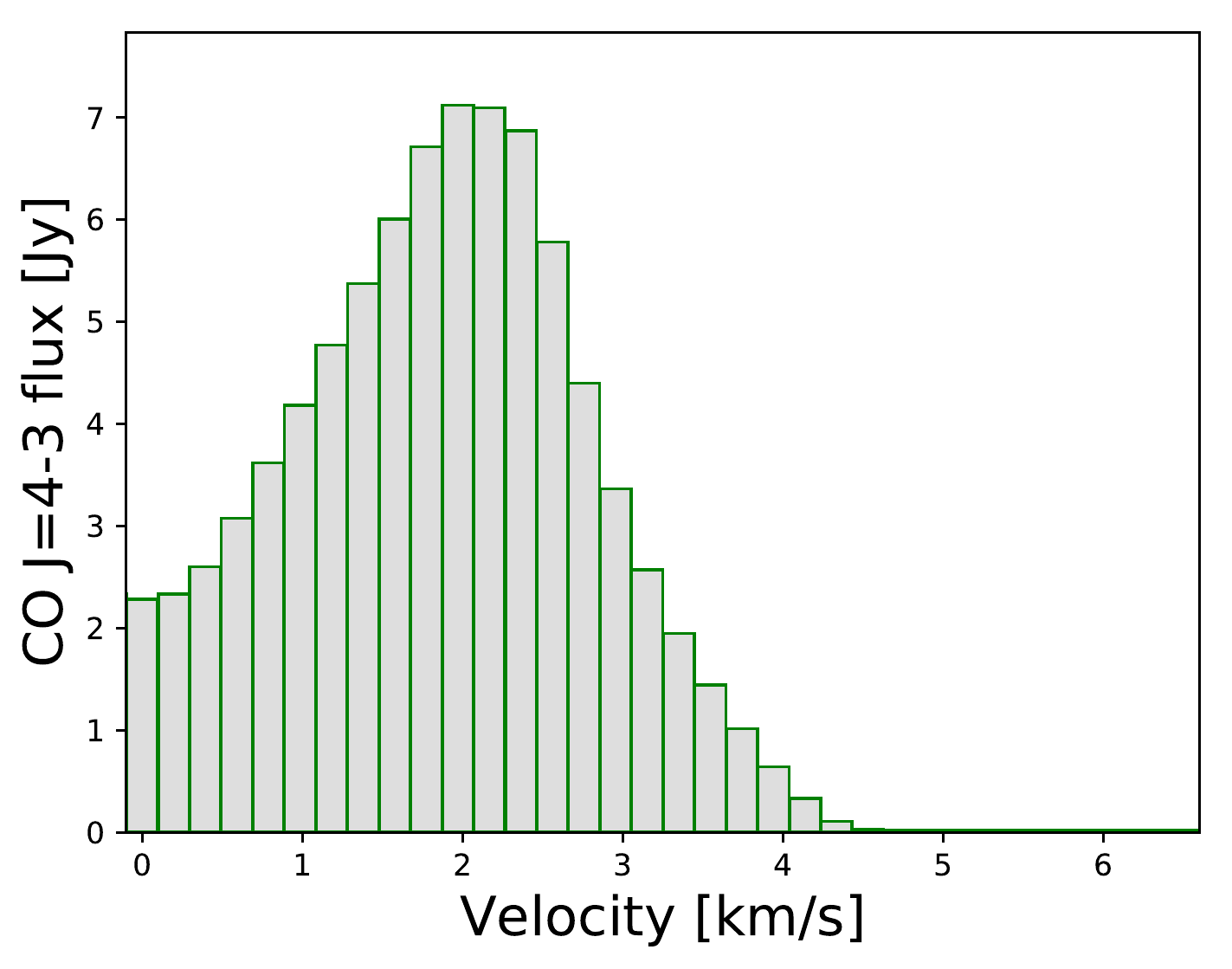}
              \subcaption{}
              \label{fig:kepler_cavity_co_line_4-3}
         \end{subfigure}
         \begin{subfigure}{.5\textwidth}
             \includegraphics[width=1\linewidth]{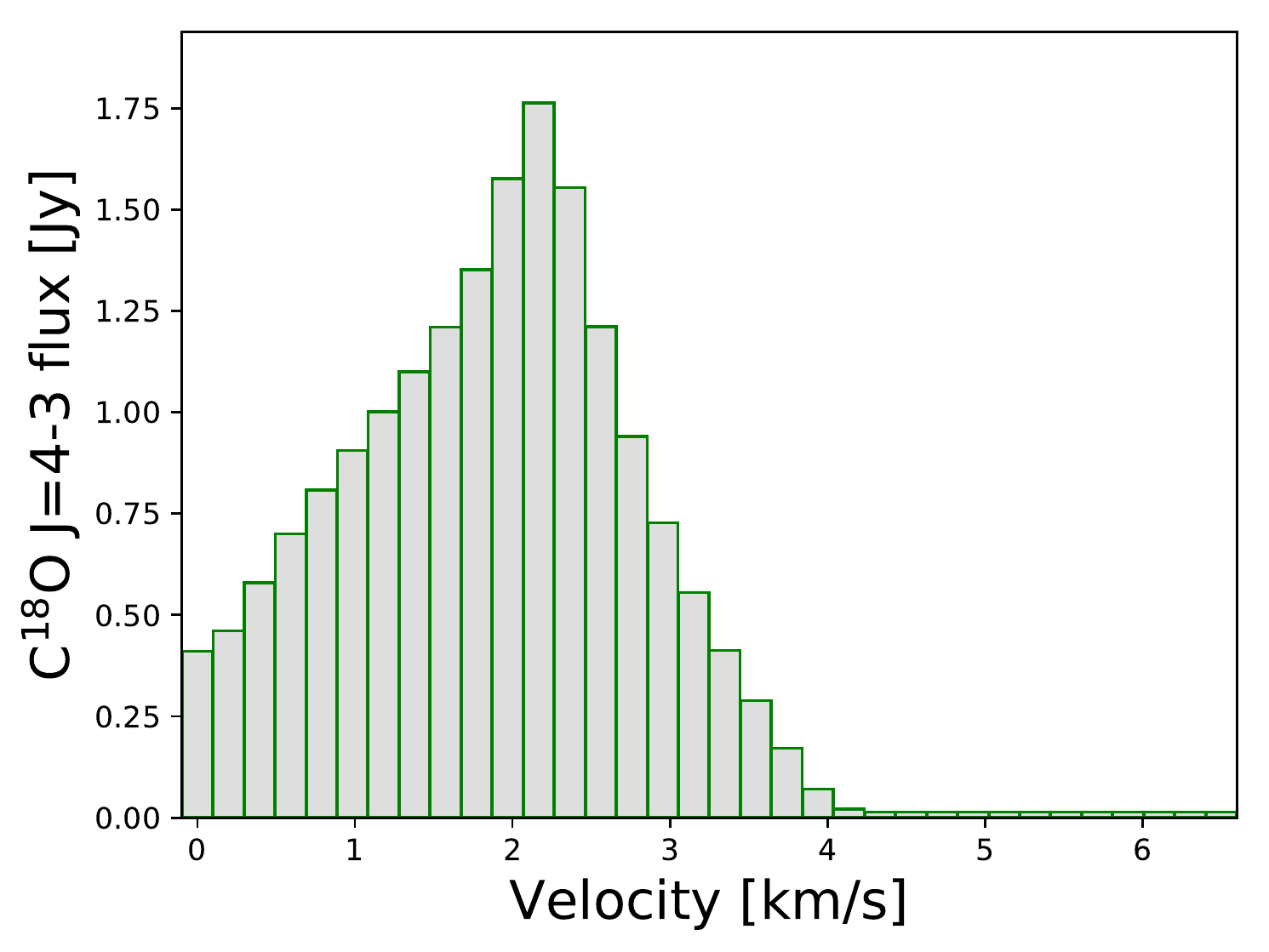}
             \subcaption{}
             \label{fig:kepler_cavity_c18o_line_4-3}
         \end{subfigure}
            \caption{   Keplerian disk  line profile (see Sect.~\ref{subsec:res_unresolved_kep} for details).
            (a) ${\rm C}{\rm O}$, $J = 4-3$.
            (b) ${\rm C}^{18}{\rm O}$, $J = 4-3$.
            (c) Keplerian disk with cavity. ${\rm C}{\rm O}$ $J = 4-3$  line emission. For details, see Sect.~\ref{subsec:res_unresolved}.
            (d) Keplerian disk with cavity. ${\rm C}^{18}{\rm O}$ $J = 4-3$  line emission. For details, see Sect.~\ref{subsec:res_unresolved}.
            }
            \label{fig:kepler_line}
      \end{figure*}

\begin{figure}[!h]
          \resizebox{\hsize}{!}{\includegraphics{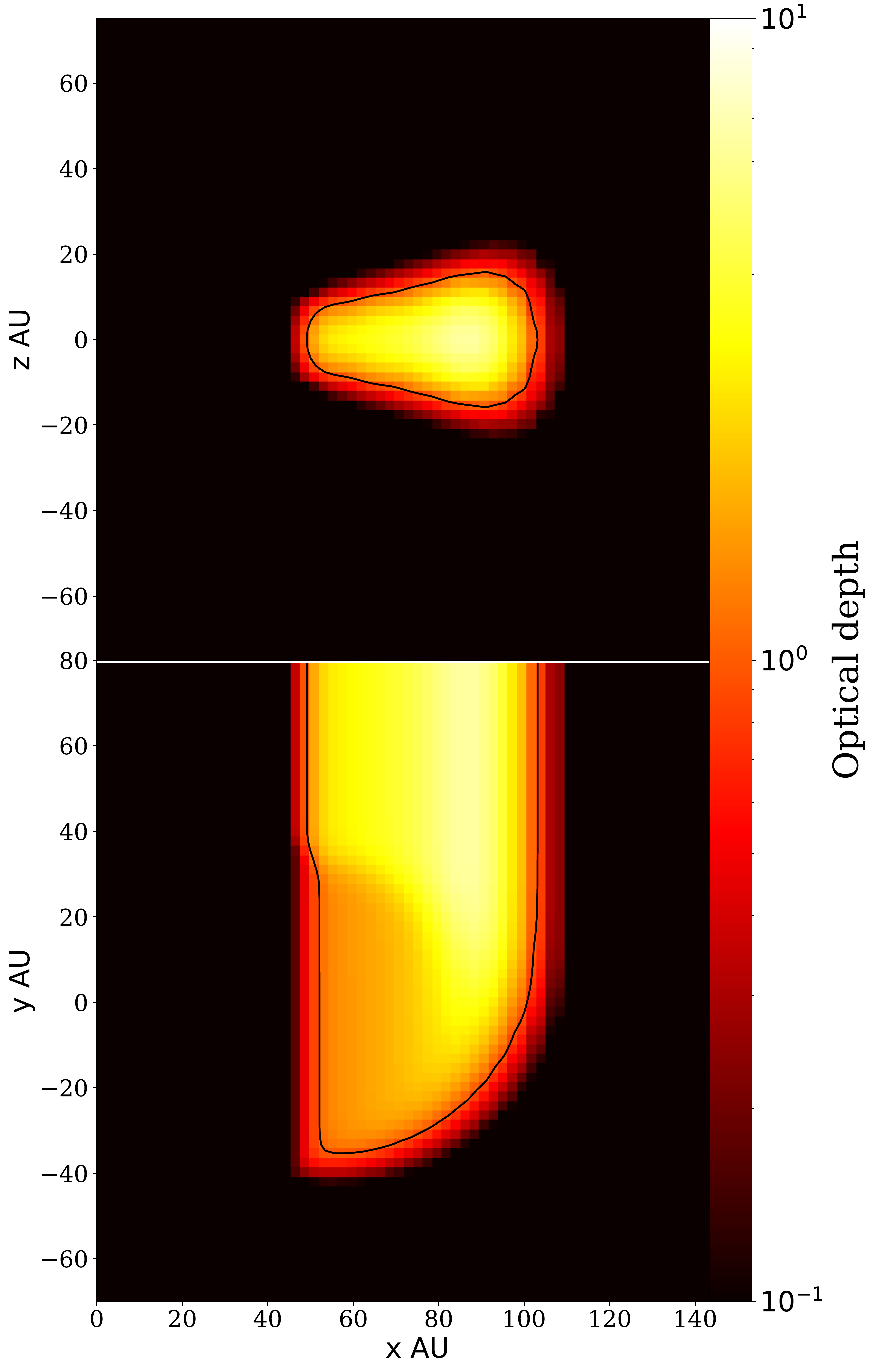}}
                   \caption{ Optical depth for the ${\rm C}^{18}{\rm O}$  $J = 3-2$ line in the Keplerian disk with cavity at $\sim 4.0$  ${\rm km/s} $ (i.e., a radius of $\sim 60$ AU).
                   The optical depth is measured (i.e., integrated) in the y direction.
                   The black contour line denotes the $\tau = 1$ boundary.}
          \label{fig:opt_dep_kepler_cavity_30_c18o_4-3}
       \end{figure}

\begin{figure}[!h]
          \resizebox{\hsize}{!}{\includegraphics{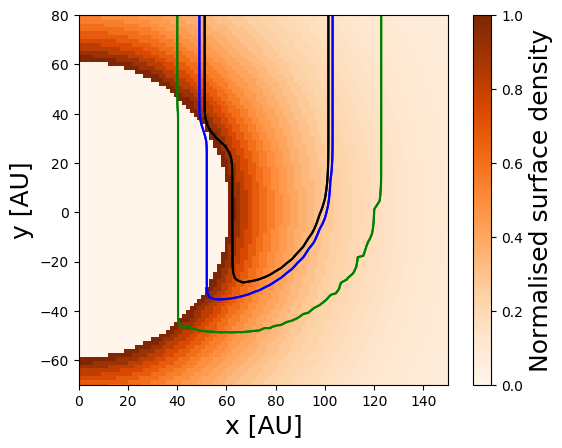}}
                   \caption{Normalized density distribution. Plotted are optical depth $\tau = 1$ boundaries for transition lines ${\rm C}{\rm O}$  $J = 4-3$ (\textit{green}), 
                   ${\rm C}^{18}{\rm O}$  $J = 4-3$ (\textit{blue}), and ${\rm C}^{18}{\rm O}$  $J = 3-2$ (\textit{black}).}
          \label{fig:density_opt_dep_kepler_cavity}
       \end{figure}

   In  Fig.~\ref{fig:velocity_channel}, we highlight the link between the spatial shape and position of individual velocity channels and the line profiles. 
   However, the shapes of channels can change depending on the optical depth of the line. The optical depth changes depending on the abundance of the 
   isotopolog and the observed transition line. Figures~\ref{fig:kepler_co_line_4-3} and~\ref{fig:kepler_c18o_line_4-3} show two examples of line profiles for 
   the CO (Fig.~\ref{fig:kepler_co_line_4-3}) and  ${\rm C}^{18}{\rm O}$ (Fig.~\ref{fig:kepler_c18o_line_4-3}) transition line $J=4-3$ for a Keplerian disk. Although the peaks are located 
   at the same velocities, the line profiles at those peaks differ significantly from each other. This can be attributed to the difference in optical depth 
   between both lines. In our models, we assumed the abundance ratio of $N({\rm CO})/N({{\rm C}^{18}{\rm O}})=500$. The peaks of the line correspond to the velocity channel reaching the 
   outer edge of the disk (Number 3 in Fig.~\ref{fig:velocity_channel}).  From this we can infer that the shape of the area in which the optical depth $\tau$ is 
   greater than unity differs significantly between isotopologs for the those channels. 
   Figure~\ref{fig:opt_dep_kepler_cavity_30_c18o_4-3} shows the optical depth of the line ${\rm C}^{18}{\rm O}$ $J=4-3$. The black contour indicates the optical depth $\tau = 1$ seen
   from the observer (y - axis). In Fig.~\ref{fig:density_opt_dep_kepler_cavity}, one can see the difference in the location of the $\tau =1$ surface between different 
   isotopologs and different lines. At this point we want to note that   although the line ${\rm C}^{18}{\rm O}$ is optically thinner compared to CO, 
   it still reaches an optical depth of $\tau > 1$ in some parts of the disk, as can be seen in Figs.~7 and \ref{fig:density_opt_dep_kepler_cavity}.

\begin{figure*}[ht]
  
         \begin{subfigure}{.5\textwidth}
             \includegraphics[width=1\linewidth]{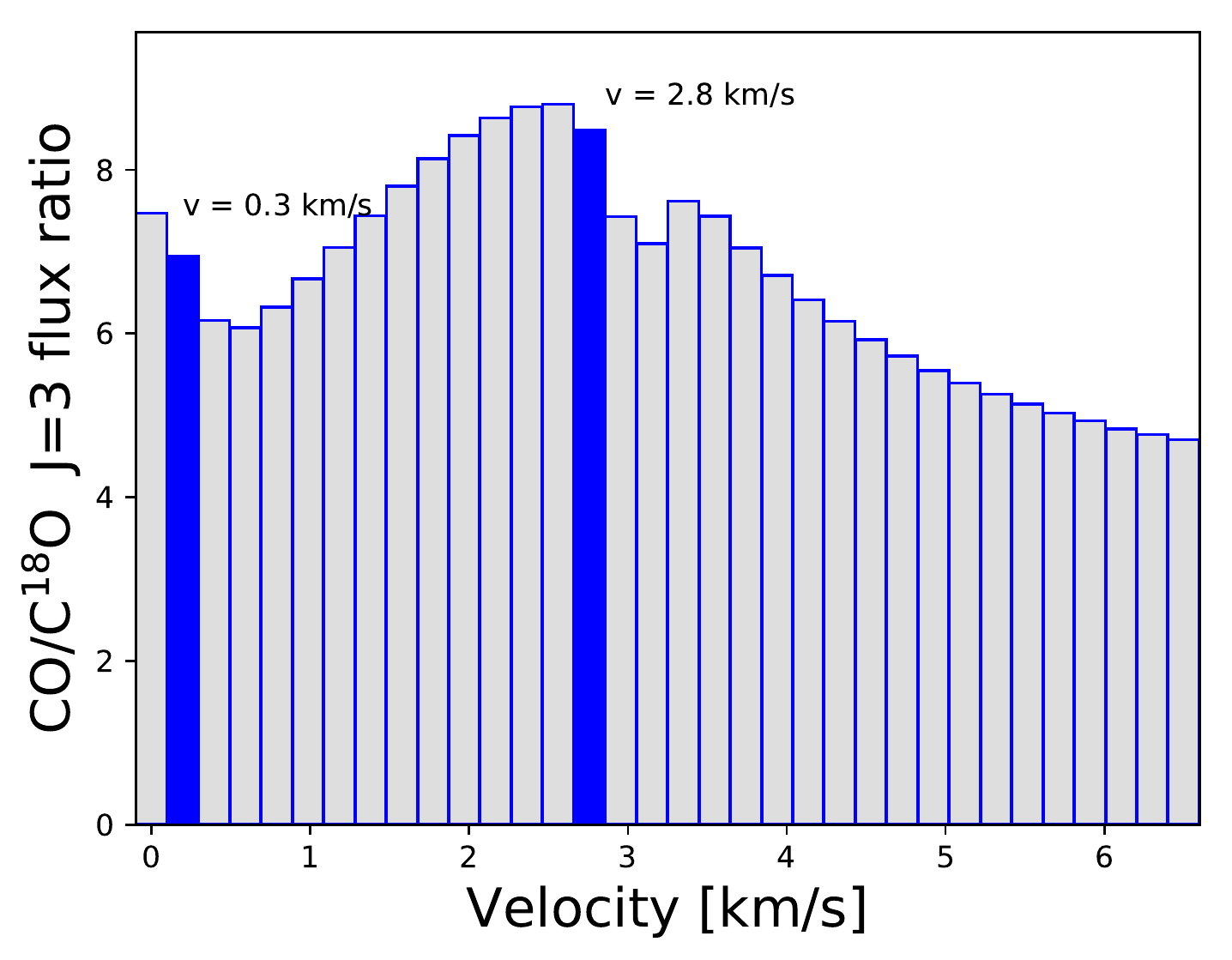}
              \subcaption{}
              \label{fig:quot_kepler_30_co-c18o_3}
         \end{subfigure}
         \begin{subfigure}{.5\textwidth}
             \includegraphics[width=1\linewidth]{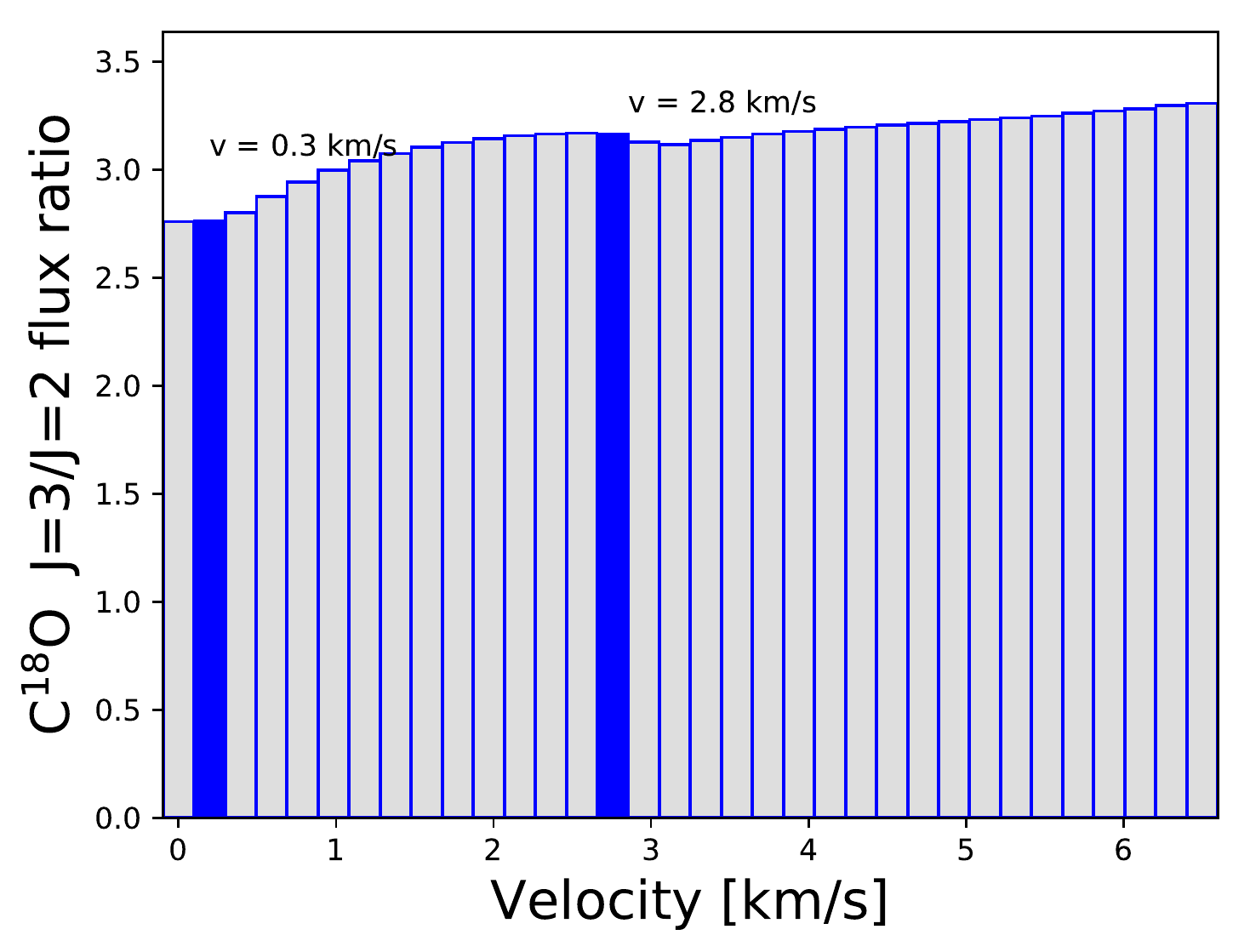}
             \subcaption{}
             \label{fig:quot_kepler_30_c18o_3-2}
         \end{subfigure}
~
         \begin{subfigure}{.5\textwidth}
             \includegraphics[width=1\linewidth]{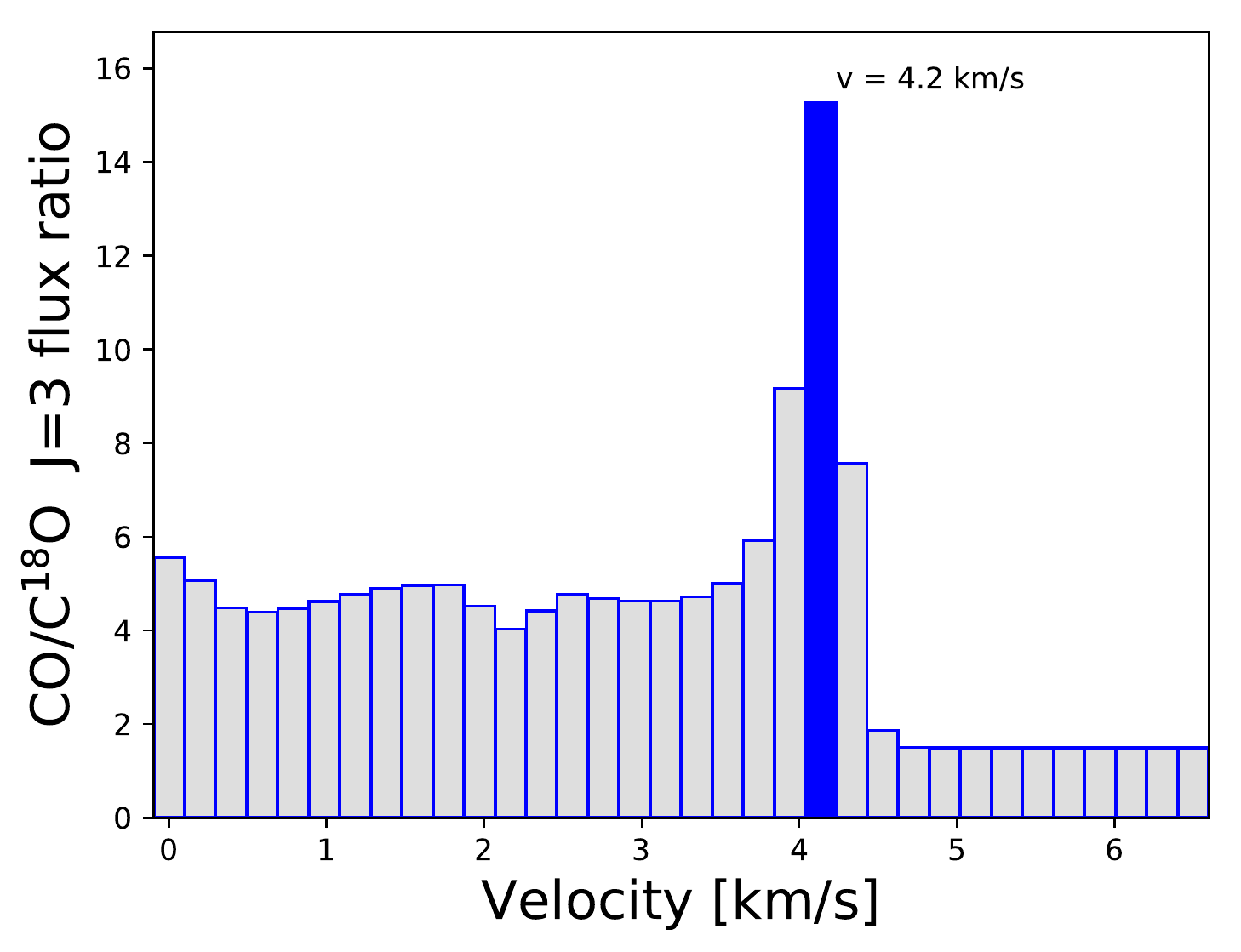}
              \subcaption{}
              \label{fig:quot_kepler_cavity_30_co-c18o_3}
         \end{subfigure}
         \begin{subfigure}{.5\textwidth}
             \includegraphics[width=1\linewidth]{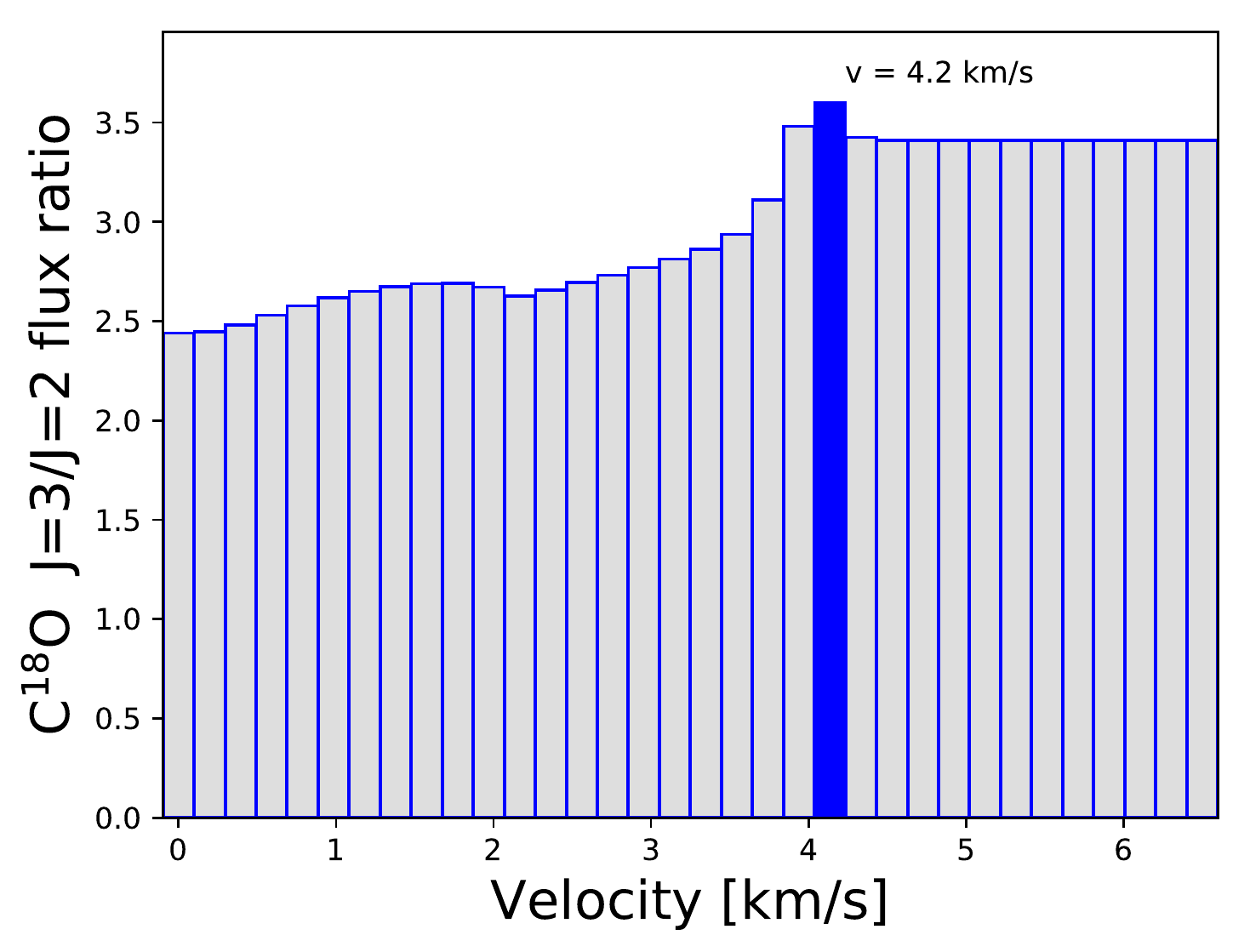}
             \subcaption{ }
             \label{fig:quot_kepler_cavity_30_c18o_3-2}
         \end{subfigure}
            \caption{   Keplerian disk: selected line emission ratio (see Sect.~\ref{subsec:res_unresolved_kep} for details).
            (a)  ${\rm C}{\rm O}$/${\rm C}^{18}{\rm O}$, $J = 4-3$. The highlighted velocity channels at 0.3 km/s and 2.8 km/s 
          indicate the features caused by self obscuring  and differences in radiating areas between the isotopologs,  respectively.
            (b) ${\rm C}^{18}{\rm O}$, $J = 4-3$/$J = 3-2$. The highlighted velocity channels at 0.3 km/s and 2.8 km/s 
          indicate the features caused by self-obscuring  and differences in radiating areas between the isotopologs,  respectively.
            (c)  ${\rm C}{\rm O}$/${\rm C}^{18}{\rm O}$  $J = 4-3$. The blue highlighted channel at 4.2 km/s indicates 
          the velocity at the cavity outer edge.
            (d) ${\rm C}^{18}{\rm O}$ $J = 4-3$/$J = 3-2$ line emission ratio. The blue highlighted channel at 4.2 km/s indicates 
          the velocity at the cavity's outer edge.
            }
            \label{fig:kepler_line_ratio}
      \end{figure*}

       Following the discussion of the individual line profiles, we now discuss the line flux ratios. Figure~\ref{fig:quot_kepler_30_co-c18o_3} shows an example of the 
       flux ratio for ${\rm C}{\rm O}$ and ${\rm C}^{18}{\rm O}$ for the line $J = 4-3$. For the Keplerian disk, the flux ratio has only two features: at 0.3 km/s
       and 2.8 km/s, respectively. Comparing the line profiles, one finds that the feature at 2.8 km/s is caused by the difference in radiating areas between the 
       isotopologs at the maximum extent of the velocity channel (Number 3 in Fig.~\ref{fig:velocity_channel}). For this reason, it can be used to determine the velocity at the 
       disk's outer edge. However, for a disk with a less sharp edge, we expect this effect to be less pronounced. The second feature at 0.3 km/s is caused by 
       the velocity channel obscuring itself (Number 5 in Fig.~\ref{fig:velocity_channel}). Once again, the difference in abundance results in a difference in size of the $\tau \geq 1$ area. 
       This, in turn, results in the self-obscuring occurring at slightly different velocities.

   For the ${\rm C}^{18}{\rm O}$   $J = 4-3$ to $J = 3-2$ line ratio (Fig.~\ref{fig:quot_kepler_30_c18o_3-2}), the features occur at the same velocity as in Fig.~\ref{fig:quot_kepler_30_co-c18o_3}. 
   The effect is smaller in both cases due to smaller differences in optical depth between the two lines (see Fig.~\ref{fig:density_opt_dep_kepler_cavity}). In the previous case, this difference 
   was the result of the difference in abundance between the two isotopologs. In this case, the difference in the optical depth is caused by the temperature gradient in the disk and the associated 
   difference in excitation of the individual levels. Those features are a geometrical effect of the velocity channels distributed in the disk and the way we observe the line emissions. 
   They will appear in some form in all graphs presented from now on and will be subsequently ignored in the main text.

    \subsubsection{Keplerian disk with cavity}
    \label{subsec:res_unresolved_kep_cav}
    
    We now discuss the case of a Keplerian disk with an inner cavity ($R = 60$ au), which is comparable in size to the depleted region in our circumbinary disk. 
    Similar to the previous subsection, we will address the individual line profiles first before continuing with the flux ratios. Figures~\ref{fig:kepler_cavity_co_line_4-3} 
    and~\ref{fig:kepler_cavity_c18o_line_4-3} show line profiles for line $J = 4-3$ for isotopologs CO and C18O.  The lack of flux at higher velocities in the center of 
    the disk is indicative of the presence of the cavity. 
    However, the velocity at which this occurs differs for both isotopologs.  Once again, we can attribute this difference to 
    the difference in optical depth. Due to a higher abundance of CO compared to ${\rm C^{18}O}$, the flux level increase is stronger for CO. This results in the maximum flux
    being reached at a higher velocity, corresponding to a smaller radius.
    Comparing this to the flux ratio shown in Fig.~\ref{fig:quot_kepler_cavity_30_co-c18o_3}, one finds that this difference results in a clearly defined maximum in the flux ratio 
    diagram.

    In the case  of the  ${\rm C}^{18}{\rm O}$ $J = 4-3$ to $J = 3-2$ (Fig.~\ref{fig:quot_kepler_cavity_30_c18o_3-2}) ratio, 
    the maximum is much less pronounced. As in the previous case, the feature is caused by the difference in optical depth between 
    the two lines and the relative level population of both levels. The lines have different excitation energies. This results in the 
    $J = 4-3$ line having a higher excitation ratio compared to the $J = 3-2$ line at the warmer, directly illuminated inner edge of the disk.
    The excitation ratio changes further inside the disk. Here the temperature is lower compared to the disk's outer layer. 
    This change manifests itself in the feature at $\sim 4.2$ $km/s$ in Fig.~\ref{fig:quot_kepler_cavity_30_c18o_3-2}.

    \subsubsection{Circumbinary disk}
    \label{subsec:res_unresolved_bin}

\begin{figure}[h!]
          \resizebox{\hsize}{!}{\includegraphics{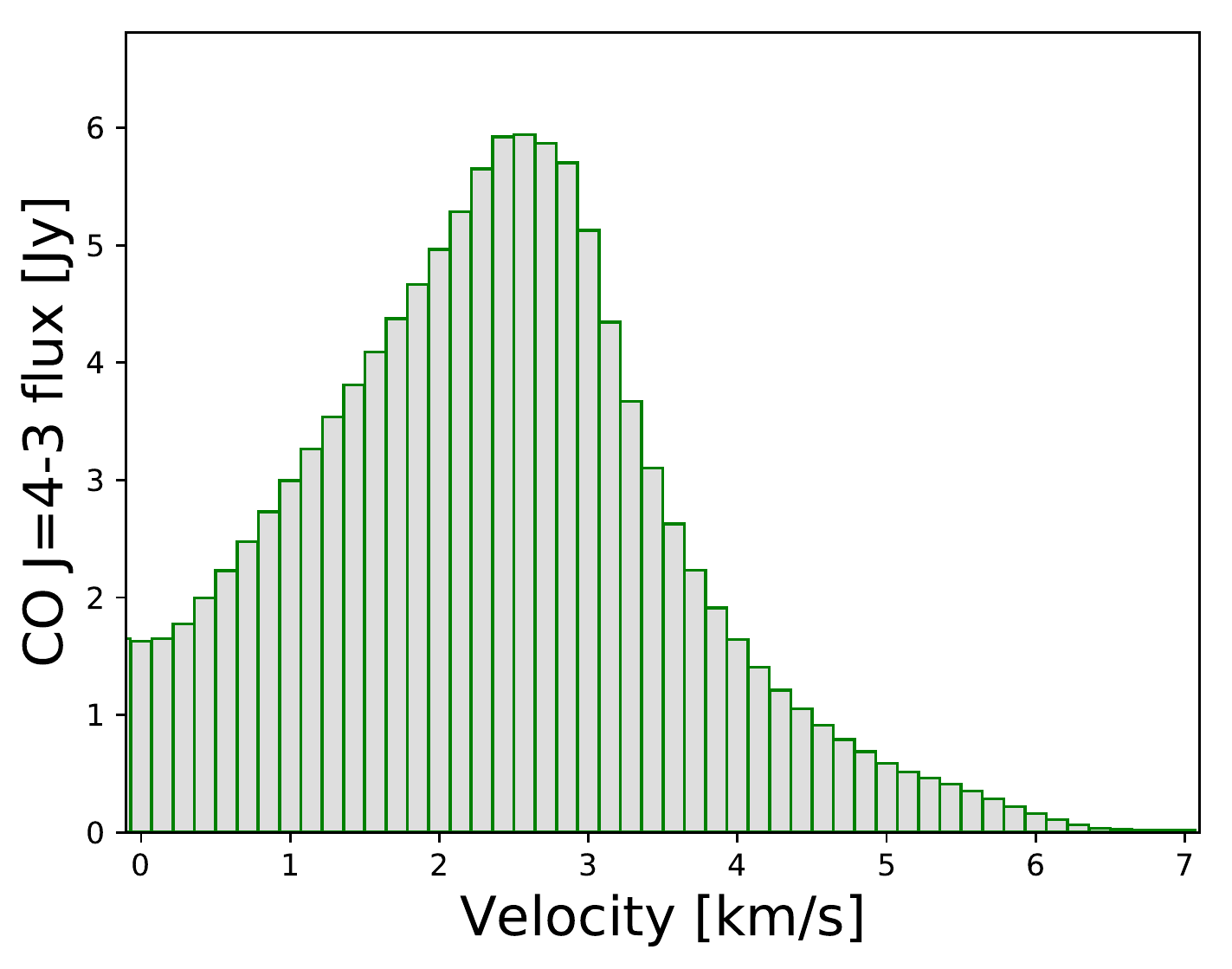}}
                   \caption{Circumbinary disk line  ${\rm C}{\rm O}$, $J = 4-3$ flux (see Sect.~\ref{subsec:res_unresolved_bin} for details).}
          \label{fig:bin_co_line_4-3}
       \end{figure}

        We begin our discussion by once again introducing an unaltered line profile. The 
        emission for ${\rm C}{\rm O}$   $J = 2-1$ for a circumbinary disk is shown in Fig.~\ref{fig:bin_co_line_4-3}. Similar to the case of a Keplerian disk with a cavity 
        (Fig.~\ref{fig:kepler_cavity_co_line_4-3}), we detect a flux deficiency for high velocities compared to the Keplerian 
        disk (Fig.~\ref{fig:kepler_co_line_4-3}). However, since the cavity in the circumbinary disk is not completely gas-free, we still detect a significant 
        flux for velocities above $4$ $km/s$. It is possible to determine the size of the inner cavity from a simple line profile, as was done by 
        \cite{Dutrey2008}. To do so, one has to perform an extensive fitting of multiple parameters simultaneously. With the flux ratios, one has an additional way of determining 
        the cavity size that does not require fitting. This can be used to further constrain the parameter range, since the resulting cavity size 
        does not depend on the other fitting parameters.

        For the circumbinary disk, we calculated all six flux ratios that can be derived from our synthetic data. 
        In particular we have three different ratios of ${\rm C}{\rm O}$/${\rm C}^{18}{\rm O}$ for each line transition and three 
        ratios for each isotopolog: $J = 4-3/3-2$ , $J = 4-3/2-1$, and $J = 3-2/2-1$.
        
\begin{figure*}[ht]
  
         \begin{subfigure}{.5\textwidth}
             \includegraphics[width=1\linewidth]{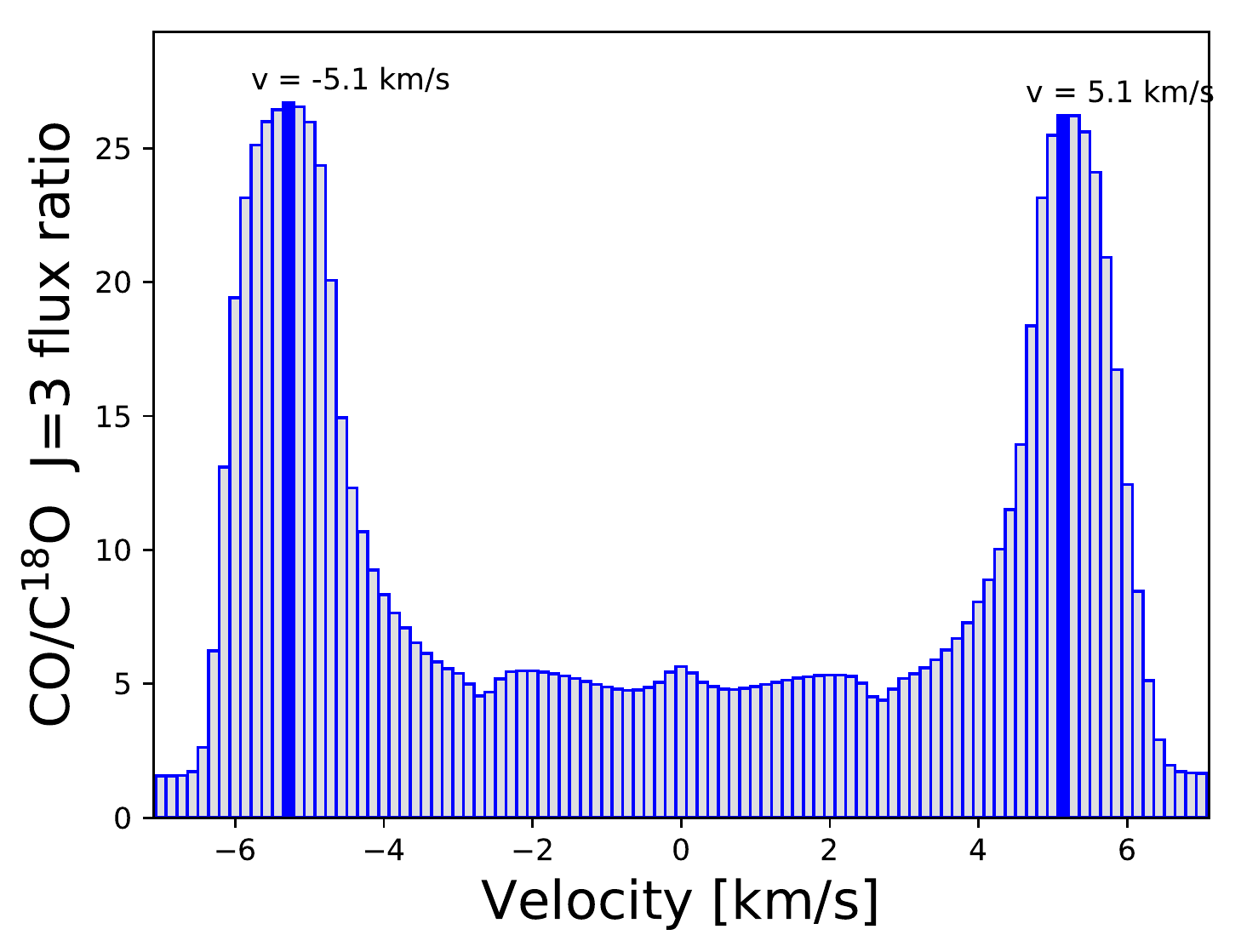}
              \subcaption{ }
              \label{fig:quot_30_co_c18o_3}
         \end{subfigure}
         \begin{subfigure}{.5\textwidth}
             \includegraphics[width=1\linewidth]{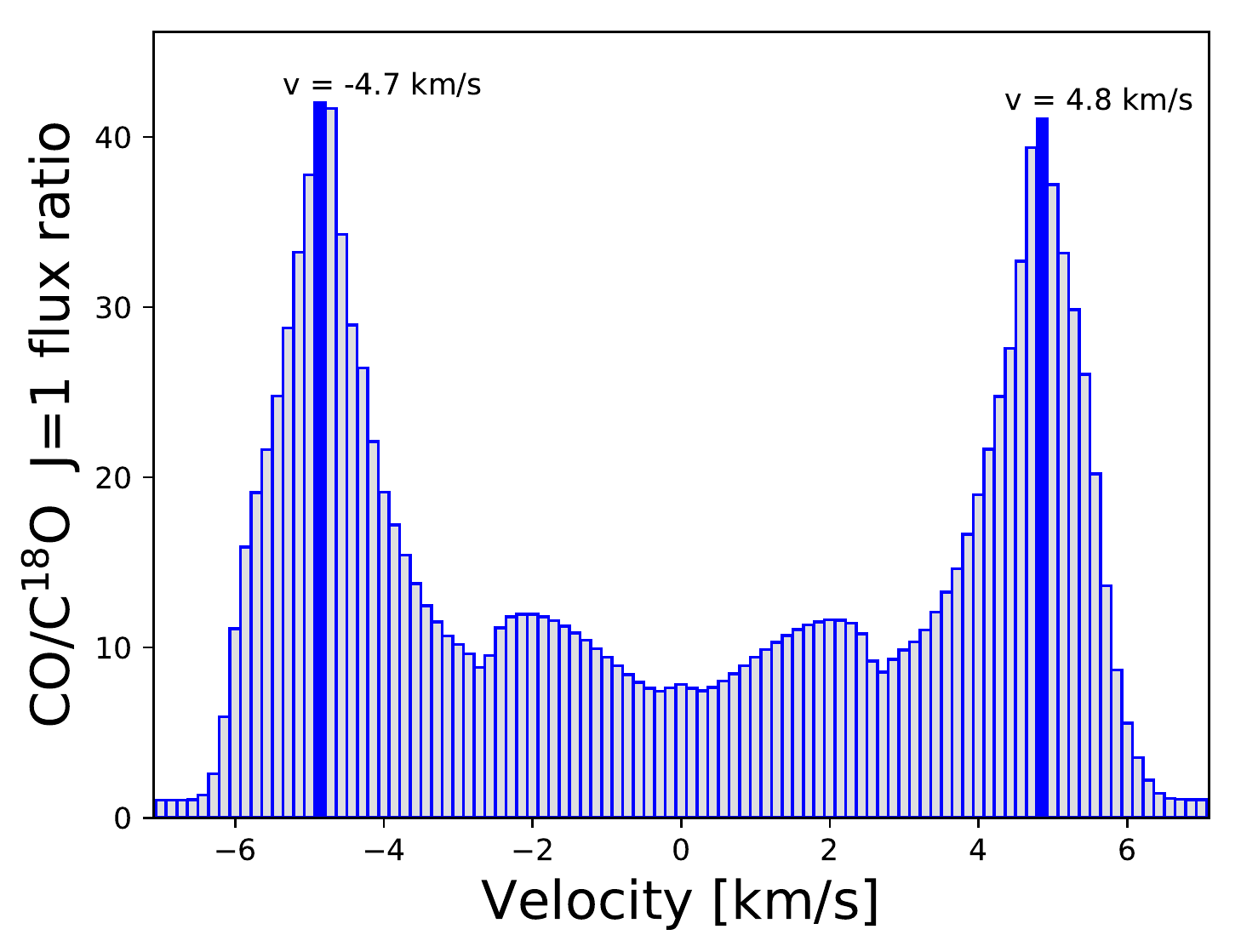}
             \subcaption{ }
             \label{fig:quot_30_co_c18o_1}
         \end{subfigure}
~
         \begin{subfigure}{.5\textwidth}
             \includegraphics[width=1\linewidth]{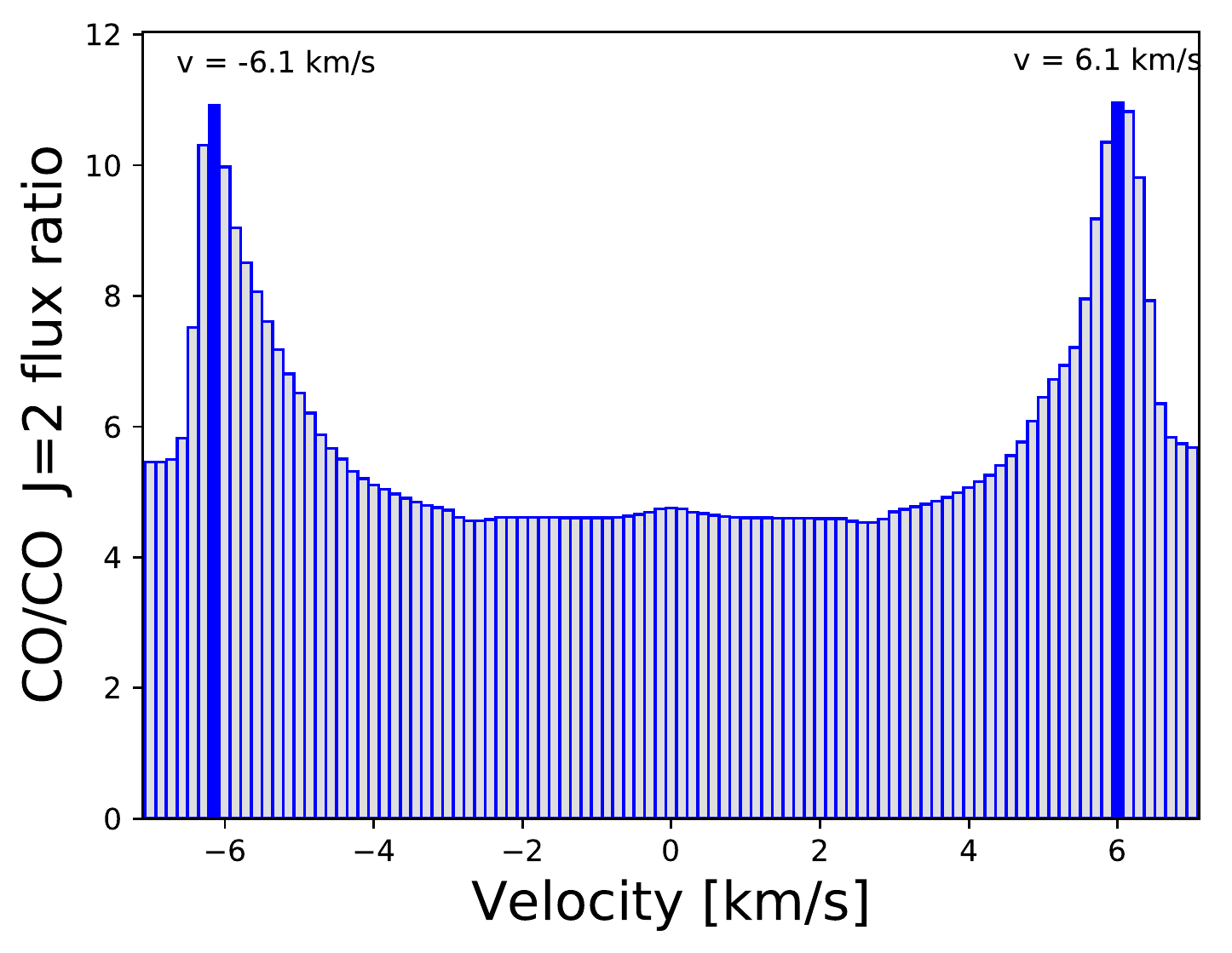}
             \subcaption{}
             \label{fig:quot_30_co}
         \end{subfigure}
         \begin{subfigure}{.5\textwidth}
             \includegraphics[width=1\linewidth]{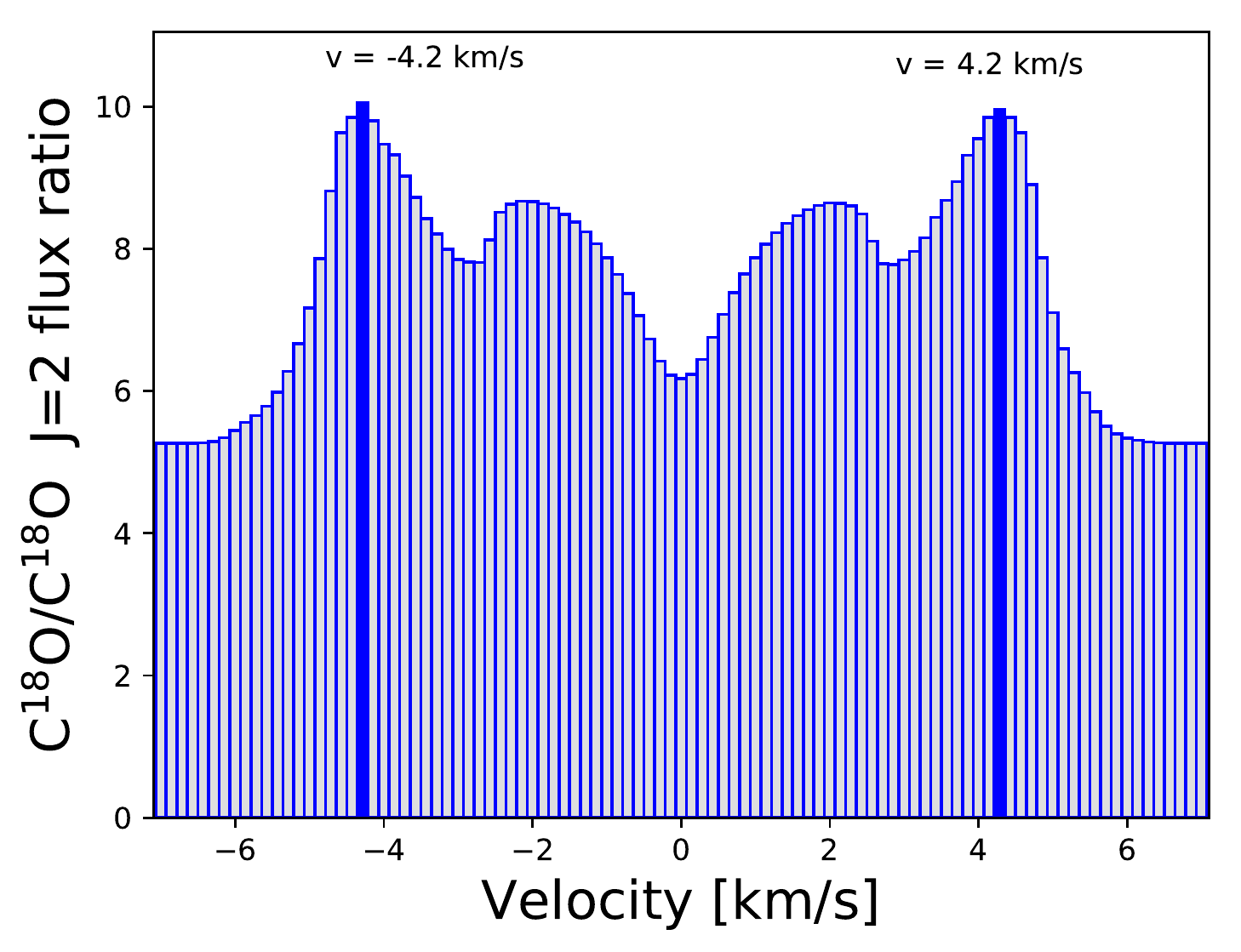}
             \subcaption{}
             \label{fig:quot_30_c18o}
         \end{subfigure}
            \caption{
            Circumbinary disk: selected line emission ratio (see Sect.~\ref{subsec:res_unresolved_bin} for details). 
            (a) ${\rm C}{\rm O}$ to ${\rm C}^{18}{\rm O}$ line $J = 4-3$.
                           The blue highlighted channel at -5.1 and 5.1 km/s indicates the velocity at the cavity's outer edge.
            (b) ${\rm C}{\rm O}$ to ${\rm C}^{18}{\rm O}$ line $J = 2-1$.
                          The blue highlighted channel at -4.7 and 4.8 km/s indicates the velocity at the cavity's outer edge.
            (c) ${\rm C}{\rm O}$ line $J = 3-2/2-1$.
                          The blue highlighted channel at -6.1 and 6.1 km/s indicates the velocity at the cavity's outer edge.
            (d) ${\rm C}^{18}{\rm O}$ line $J = 3-2/2-1$.
                          The blue highlighted channel at -4.2 and 4.2 km/s indicates the velocity at the cavity's outer edge.
                         }
            \label{fig:bin_line_ratio}
      \end{figure*}

        Comparing the flux ratio for different isotopologs in Fig.~\ref{fig:quot_30_co_c18o_3} to the example of a Keplerian disk 
        with a cavity (Fig.~\ref{fig:quot_kepler_30_co-c18o_3}), one notices a number of major differences. The first is the asymmetry between 
        the right and the left side of the graph. This is caused by the asymmetry in the density  of the hydrodynamical distribution. 
        This allows for an independent investigation of both disk sides.

   We now discuss the flux ratio between different transition lines for the same isotopolog. The feature corresponding to the 
   inner disk rim appears at a higher velocity of $v_{\rm max} = 6.1$ ${\rm km/s}$ in the case of ${\rm C}{\rm O}$  (Fig.~\ref{fig:quot_30_co})
   compared to $v_{\rm max} = 4.2$ ${\rm km/s}$ for ${\rm C}^{18}{\rm O}$  (Fig.~\ref{fig:quot_30_c18o}); this is in agreement with 
   the result we derived from the simulated spatially resolved observations (see Sect.~\ref{subsec:res_resolved_bin} and Fig.~\ref{fig:density_circle} for comparison).

   \subsubsection{Observability}
   
   Spatially unresolved observations do not have the constraint of minimum-required spatial resolution, and they reduce the required  sensitivity for the method presented. 
   However, we still have to evaluate whether the sensitivity of the current instruments, ALMA in particular, is high enough to observe the studied features, in this case the cavity in the center of the disk. 
   The peak fluxes in the synthetic images presented range between $0.435$ Jy  for ${\rm C}^{18}{\rm O}$ line $J = 2-1$ 
   and $40.9$ Jy  for ${\rm C}{\rm O}$ line $J = 4-3$. However, we are not interested in the regions that contribute to the line maxima, but rather the wings of the line. 
   At these wings, the flux reaches values of, at most, an order of magnitude below the line maximum.
   
\begin{figure}[h!]
          \centering
                 \resizebox{\hsize}{!}{%
                                      \includegraphics[width=0.6\columnwidth]{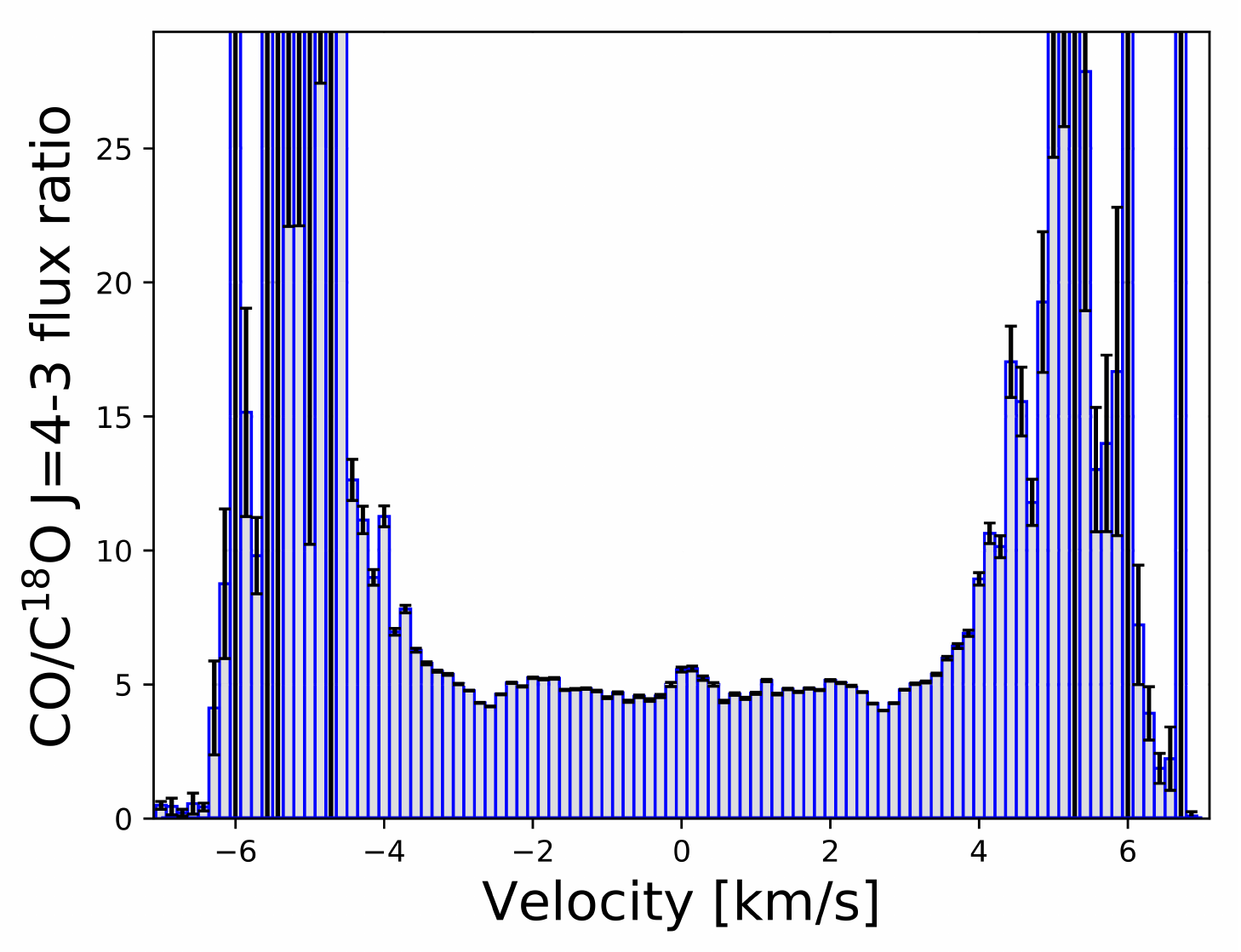}%
                                     }
          \caption{Synthetic observation of a circumbinary disk ${\rm C}{\rm O}/{\rm C}^{18}{\rm O}$ line $J = 4-3$ emission ratio.  
          Configuration 1 of the ALMA Observational Cycle 7 is used with the integration time of 2 h and assumed distance of 140 pc.
          The black error bars indicate a 1$\sigma$ uncertainty.}
          \label{fig:quot_30_co_c18_3_casa}
        \end{figure}

   Since we are not interested in a high spatial resolution, we chose the ALMA antenna configuration with the smallest baseline possible (i.e., configuration Number 1 from
 Observational Cycle 7). This results in a resolution of 3.38 as, 1.47 as, and 0.98 as for lines $J = 1-2$, $J = 2-3$, and $J = 3-4$, respectively. The difference in transition frequency 
   between both isotopologs is small enough for the resolution to be considered the same in both cases. At an assumed object distance of 140 pc and disk diameter of 400 AU, the disk has an angular 
   extent of 2.85 as. However, at a spectral resolution of 141 m/s, the area contributing to each individual channel is much smaller. This makes it possible to cover the contributing regions with a single beam 
   for all three lines. As was the case with the synthetic lines in the previous subsection, we integrated the flux over the spatial dimensions for each velocity channel.  We simulated
   an ALMA observation of 2 hours using the ALMA simulation tool kit  \texttt{CASA} \citep{McMullin_2007} for three different lines for  both isotopologs. 
   For this, we assumed the default value for the precipitable water vapor of 0.5 mm and an ambient temperature of 269.0 K. By comparing the integrated flux 
   values before and after applying \texttt{CASA}, we can assure the flux conservation up to the noise level. Figure~\ref{fig:quot_30_co_c18_3_casa} shows an example of a 
   line ratio observed with ALMA. In this image, one can discern the general form of a flux ratio before it is processed with \texttt{CASA} (compare to Fig.~\ref{fig:quot_30_co_c18o_3}). 
   The ratios coincide well in channels with higher fluxes. Since it is not possible to determine the location of the ratio maxima from Fig.~\ref{fig:quot_30_co_c18_3_casa}, 
   it is not possible to determine the radius of the cavity.
   However, one can clearly see the increase in flux ratios around $\pm$ 4.5 km/s. With this we can can still give an upper boundary for the cavity radius.
   
   While we assumed typical disk properties in our study, the presented method becomes much more 
   powerful if applied to a disk with higher apparent brightness, such as the Flying Saucer in \citep{Dutrey_2017}.
   Their observations were performed under conditions similar to those described here. The two major differences were a significantly higher maximal surface
   brightness of the Flying Saucer of about a factor of ten and a larger CO disk extent of 5.5 (compared to the 2.85 used here).

   \subsubsection{Spatial distribution}
   
\begin{figure}[h!]
          \resizebox{\hsize}{!}{\includegraphics{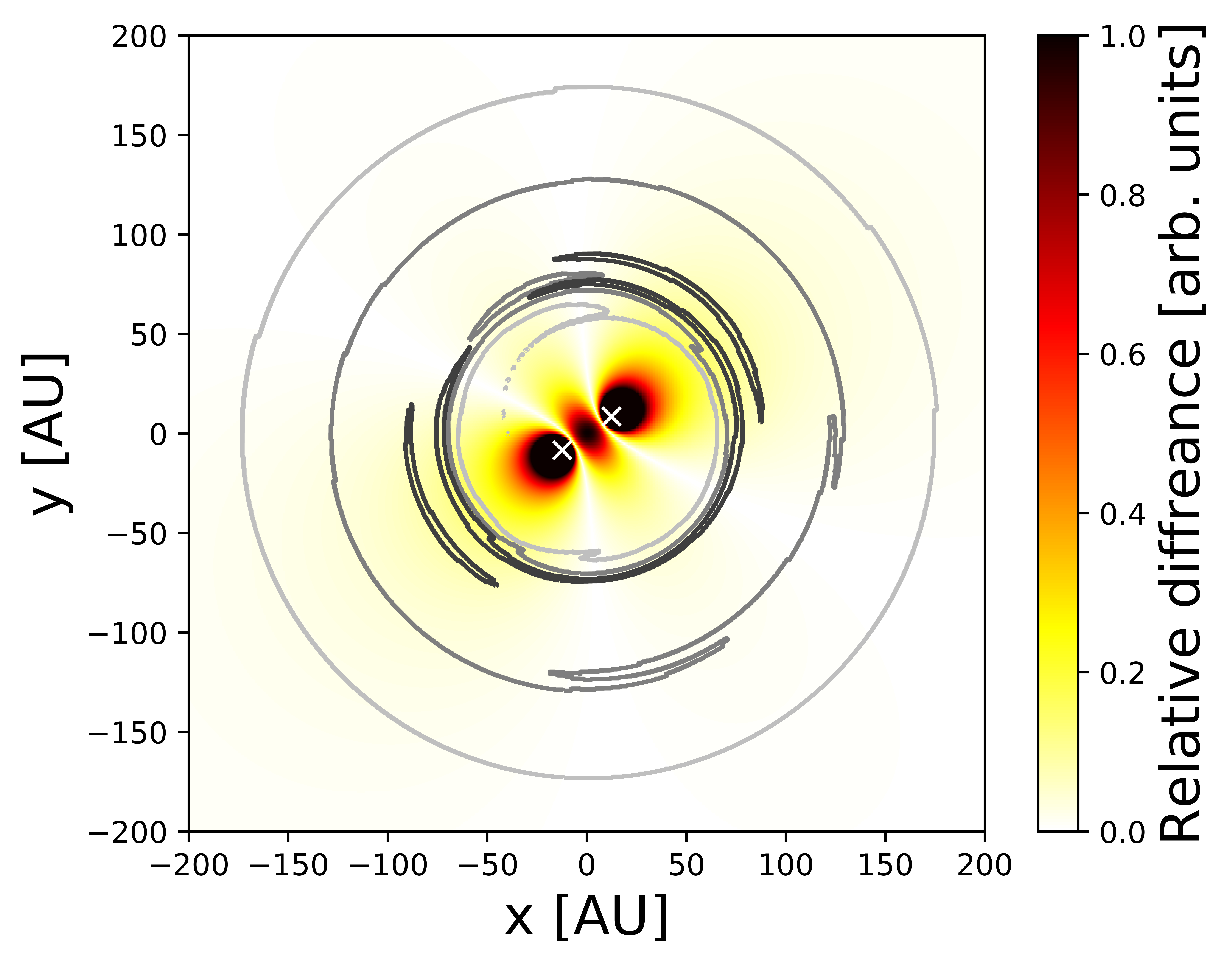}}
          \caption{The absolute value of the difference between the gravitational potential of the binary and a single star with the combined mass divided by the single star potential.}
          \label{fig:grav_pot}
        \end{figure}

   So far we have only considered the correlation between the location of the disk's inner edge  and the velocity  $v$. 
   Beyond this it is possible to match each velocity channel to a specific volume inside the disk. For this we need the velocity 
   field of the circumbinary disk. Assuming that it is mostly governed by the balance between the gravitational and centrifugal forces, 
   it is sufficient to know the binary masses $M_1$ and $M_2$ and their positions (in the x-y plane according to our reference coordinate system). 
   It is possible to use the reduced mass of the binary $\mu = (M_1\cdot M_2)/(M_1 + M_2)$ and presume the velocity field to be near-Keplerian. 
   The reduced mass can be estimated from the radius of  the outer edge of the disk and the corresponding gas velocity. The gas velocity can be derived with the 
   associated feature in the line ratio discussed in Sect.~\ref{subsec:res_unresolved} or by measuring the total line width for each line. 
   However, as the binary and single star velocity profiles differ, this leads to errors in determining the spatial location of the ratio features.
   In Fig.~\ref{fig:grav_pot}, the absolute value of the 
   difference between the gravitational potential of the binary and a single star with the combined mass divided by the single star potential is shown. 
   A contour plot of the midplane gas density is superimposed over the relative difference plot. The white X symbols mark the binary positions.  
   At the inner rim of the disk, the region we are most interested in, the difference in the potentials still amounts to 20 percent and decreases further in the disk.

\begin{figure}[h!]
          \resizebox{\hsize}{!}{\includegraphics{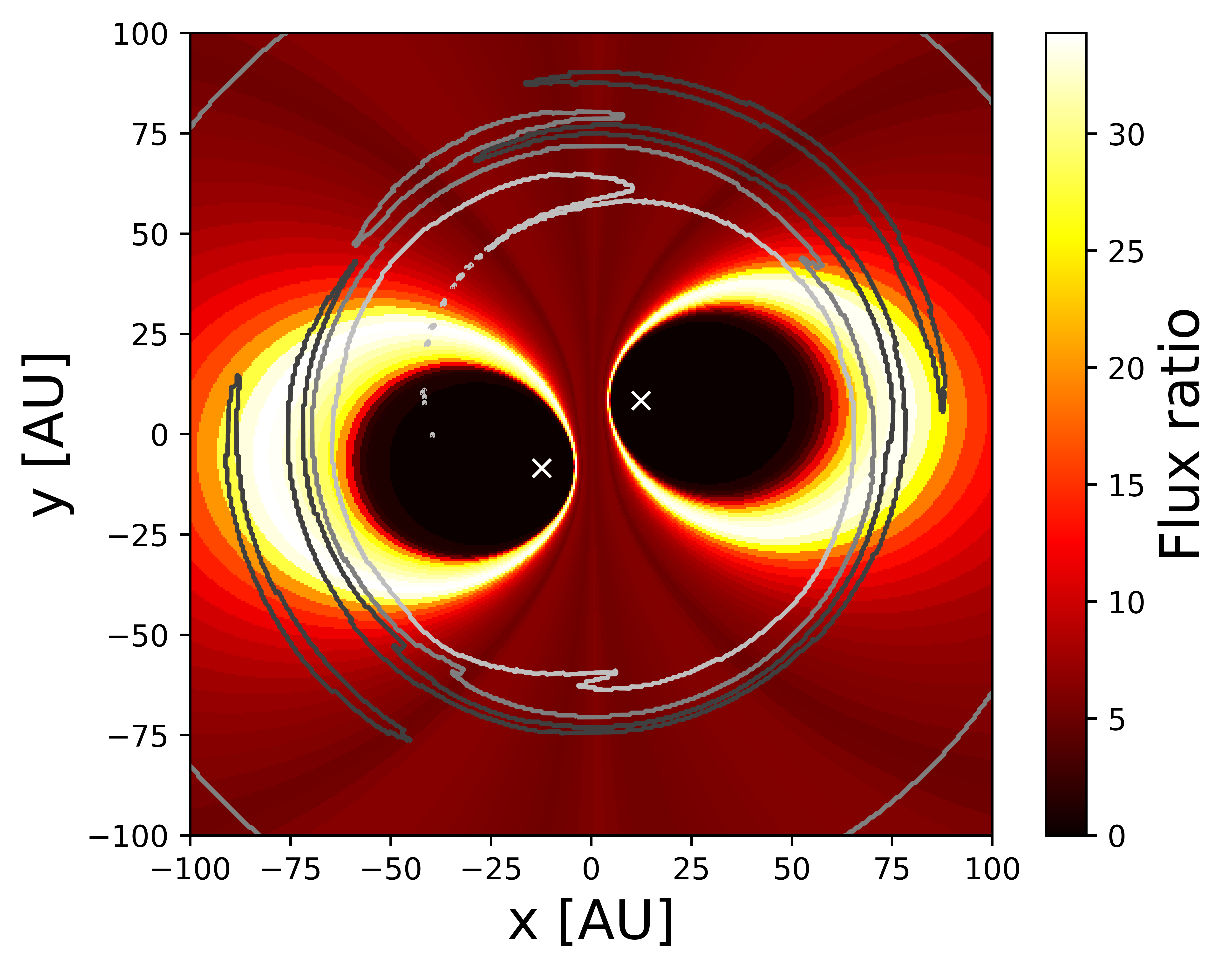}}
          \caption{Line emission ratio for ${\rm C}{\rm O}$ and ${\rm C}^{18}{\rm O}$ $J = 3-2$. The contour plot shows the surface density profile 
          of the circumbinary disk. The white X symbols denote the binary positions.}
          \label{fig:co-c18o___line_nr:_2_xy-plane}
        \end{figure}

   Figure~\ref{fig:co-c18o___line_nr:_2_xy-plane} shows a reconstruction of 
   the inner disk structure using the line flux ratio. For this we used the binary mass and binary position data from our simulations. The binary 
   positions are once again marked with white X symbols. The binary positions were used to calculate the velocity field of the system. The velocity  field was binned 
   in the same manner as the velocity channels. Subsequently, we assigned the area of the disk associated with a specific channel the line ratio derived in that channel. 
   The midplane density distribution was overlaid as a contour plot.
   In Fig.~\ref{fig:co-c18o___line_nr:_2_xy-plane}, one finds that the outer edge of the feature seen in ratio maps, corresponding to higher orbital velocity, 
   coincides with the inner rim of the disk. Furthermore, one finds that the inner edge of the feature coincides with the second maximum
   of the density  oscillation. The inner edge of the feature is the point at which both lines become optically thick.

    \subsubsection{Model scaling}
    
    In this subsection, we discuss the influence of some of the more important model parameters on the studied features. We then, in turn, discuss the feasibility of deriving them from the observations.

    The two main features discussed here are the inner cavity and the density waves. The size of the cavity is determined by the location of the last  outer Lindblad resonance.
    In the case of a binary of equal mass, it is only dependent on the binary separation $a$ (see~\citealt{Artymowicz_Lubow_1994}; Eq. 2.2). 
    A larger binary semimajor axis $a$ would result in a larger cavity. This, in turn, would make it easier to detect. The density waves propagate outward
    from the disk's inner edge with the sound speed $c_s$ as the phase velocity. The wavelength of the oscillation is therefore  dependent on the binary period $P$ (see Sect. 3.1 of \cite{Avramenko_Wolf_2017} ). 
    The binary period can be influenced by changing the mass as well as by changing the semimajor axis. Thus,
    decreasing the binary separation would result in smaller and therefore less detectable density waves and cavities. Additionally, a smaller cavity would lead to higher 
    temperatures at the disk's inner edge. For that reason, one would have to turn to molecular lines with higher excitation energies. 
    Changing the binary mass, and therefore the temperature, would, for once, have a similar effect: Here we would expect a modified temperature at the inner rim.  On the other side,
    though it does not change the cavity size, the binary mass has a significant impact on the wavelength of the density waves.
    A lower binary mass results in an increased wavelength of the density wave, making it easier to detect this phenomenon.
    
    Regarding the disk, the two parameters of interest are the disk mass and the abundance of the used isotopologs. We do not expect the disk mass 
    to have a major impact on the applicability of the method. In the case of a spatially resolved observation, one detects the density gradient while the spatially unresolved case locates the 
    optical depth $\tau =1$ surface for different lines. As long as the disk mass remains large enough for the disk to remain optically thick, it is still possible to locate the density gradient.
    The isotopolog abundances have the potential to alter the results, since the location of the $\tau =1$ surface would be changed for each molecule individually. 
    This would change the shape of the line ratio graphs  significantly.

    For our study, we assumed the binary system to have a distance of 140 pc, motivated by the average distance to the star forming regions in Taurus. As was shown in the previous
    section, the sensitivity of ALMA is not sufficient for the synthetic observations considered here. Decreasing the distance would result in higher flux levels for all lines 
    (geometrical thinning). This, in turn, would result in an improved observability of the considered features in spatially resolved and unresolved cases.

   \section{Conclusions}
   
   The goal of this study was to investigate the feasibility of constraining disk parameters with line observations of a circumbinary disk with 
   an edge-on orientation. In order to do that, we conducted 2D hydrodynamic simulations. 
   The resulting density distribution as well as the associated velocity field were used as inputs for subsequent
    radiative transfer simulations. In the
    first step, the 3D temperature distribution was calculated, which, in turn, was used to calculate the 
    level populations and resulting line emission maps. 
    
    The study itself was split into two parts. In the first part we considered the case of spatially resolved observations. 
    We find that, despite the edge-on orientation, the Doppler effect allows us to distinguish all  
     characteristic disk regions expected of the considered circumbinary disk model. In the case of a Keplerian disk with a cavity, the cavity size was determined.
     For the Keplerian disk with a gap, the gap's inner and outer radii were derived. Using the same method, the wavelength of the density oscillation
     present in a disk can be extracted. Those features were subsequently shown to be present in synthetic circumbinary disk observations. 
     In particular, we find clear indicators for the presence of a
    central cavity, accretion arms, and the density wave structure. 
    The derived cavity sizes vary with the considered isotopolog and transition line. We have attributed it to the 
    density gradient at the disk's inner edge.

    Because of the high sensitivity requirements for a spatially resolved observation, we also investigated the potential of deriving constraints on the circumbinary disk structures from  
      spatially unresolved observations. For this we derived the line flux ratios for different isotopolog-transition lines.
    In the reference case of a Keplerian disk with a central cavity, we find that it is still possible 
    to derive the orbital velocity at the disk's inner edge from the flux ratio maps. As was the case for spatially resolved images, this velocity varies with isotopologs and 
    transition lines. Subsequently, we identified the same features in flux ratio maps for the circumbinary disk. 
    With regard to the observability of those features with the current instruments, we find that the sensitivity of
    ALMA is only sufficient to give an upper boundary for the cavity size.
    By assuming binary masses and positions, we connected the derived velocity with a specific radius. 
    Furthermore, this allowed us to infer information about the gas distribution around the disk's inner edge.

   \begin{acknowledgements}
   We thank all the members of the Astrophysics Department Kiel for helpful discussions and remarks, as well as the 
   language corrections. 
   This study was funded by the German Science Foundation (DFG), grant: WO 857/12-1.
   \end{acknowledgements}

%
%
\bibliographystyle{aa} 
\bibliography{35610corr} 

\end{document}